\documentclass[pra,onecolumn,superscriptaddress,aps]{revtex4}
\usepackage{amsfonts}
\usepackage{amssymb,latexsym,amsmath}
\usepackage{graphicx}
\usepackage{times}
\usepackage[caption=false]{subfig}
\usepackage[usenames]{color}

\begin{document}

\title{Solitary Waves of the Two-Dimensional 
Camassa-Holm--Nonlinear Schr{\"o}dinger Equation}


\author{C. B. Ward}
\thanks{Corresponding Author: ward@math.umass.edu}
\affiliation{Department of Mathematics and Statistics, University of
	Massachusetts, Amherst MA 01003-4515, USA}

\author{I. K. Mylonas}
\affiliation{Department of Mathematics and Statistics, University of
	Massachusetts, Amherst MA 01003-4515, USA}

\affiliation{Department of Mechanical Engineering, Faculty of Engineering Aristotle
	University of Thessaloniki, 54124 Thessaloniki, Greece}

\author{P. G. Kevrekidis}
\affiliation{Department of Mathematics and Statistics, University of
	Massachusetts, Amherst MA 01003-4515, USA}

\author{D. J. Frantzeskakis}
\affiliation{Department of Physics, University of Athens,
Panepistimiopolis, Zografos, Athens 15784, Greece}

\begin{abstract}

  In this work, we study solitary waves in
  a  $(2+1)$-dimensional   variant of the defocusing
nonlinear Schr\"{o}dinger (NLS) equation, the so-called Camassa-Holm NLS (CH-NLS) equation. 
We use asymptotic multiscale expansion methods to reduce this model to a Kadomtsev--Petviashvili 
(KP) equation. The KP model includes both the KP-I and KP-II versions, which possess line 
and lump soliton solutions. Using KP solitons, we construct approximate
solitary wave solutions 
on top of the stable continuous-wave solution of the original CH-NLS model, which are 
found to be of both the dark and anti-dark type. We also use direct numerical simulations to 
investigate the validity of the approximate solutions, study their evolution, as well as 
their head-on collisions. 

\end{abstract}

\maketitle
        
\section{Introduction}

It is well known that numerous
physically relevant nonlinear evolution equations,  
are completely integrable via the inverse scattering transform, and 
possess soliton solutions \cite{ablo91,ablo2011}. In particular, universal equations like the 
nonlinear Schr{\"o}dinger (NLS), the Korteweg-de Vries (KdV) --as well as its two-dimensional 
generalization, the Kadomtsev--Petviashvilli (KP) equation--, the
sine-Gordon equation and others, are ubiquitous 
in numerous physical settings, such as water waves, plasmas,
mechanical systems, nonlinear optics,
Bose-Einstein condensates, and so on \cite{ablo2011,infeld,rem,rcg:BEC_BOOK}. 
Interestingly, these soliton equations can be connected --i.e., reduced one to the other-- 
via multiscale expansion methods \cite{zakharov86}. Such connections have  
proved to be extremely useful in the study of non-integrable models: for instance, 
in the context of optics, the reduction of certain perturbed defocusing NLS equations 
to the KdV model allowed not only for the derivation of approximate dark
(gray) soliton 
solutions, but also the prediction of novel structures, the anti-dark solitons 
having the form of humps (instead of dips) on top of a continuous-wave background 
\cite{kivshar90,kivshar91,m1}. Importantly, relevant studies have also been performed 
in two-dimensional (2D) settings, leading to the prediction of structures 
such as ring solitons \cite{r1,r2,r3,r4,r5,r6}, as well as line solitons and lumps 
satisfying effective KP equations \cite{2d1,hec,2d2,gino1,2d3}.

On the other hand, recently, there has been an interest on the study of deformations 
of integrable equations, which can give rise to novel interesting models
in their own right. 
For instance, the deformation of the KdV equation leads to the 
Camassa-Holm (CH) equation (which is also
an integrable model originally derived in 
the context of water waves \cite{ch}), while the deformation of the NLS equation 
results in a variant that has been referred to
as the Camassa-Holm--Nonlinear Schr{\"o}dinger (CH-NLS) equation 
\cite{Arnaudon,Arnaudon2}. The focusing version of the CH-NLS model was studied 
in Ref.~\cite{Arnaudon2}, both analytically and numerically, and its bright soliton 
solutions, as well as their dynamics and interactions were explored.
Notice that hereafter the term soliton will be used even when the model
is not necessarily integrable, referring to solitary wave structures
in such a case.
Furthermore, in 
the recent work \cite{2017} similar investigations, but for the defocusing CH-NLS 
model, were performed; in fact, the methodology used in Ref.~\cite{2017} was relying 
on the reduction of the CH-NLS model to the KdV equation, which allowed for the 
derivation of dark and anti-dark soliton solutions. Nevertheless, apart from the 
above mentioned works, to the best of our knowledge, there exist no studies 
devoted to the 2D version of the CH-NLS equation.

It is the purpose of this work to contribute to this direction by studying the 
$(2+1)$-dimensional defocusing CH-NLS equation. Here, we will adopt methods of 
multiscale expansions to eventually reduce the considered model to a KP equation; 
this allows us to construct approximate soliton solutions, having the form of 
line solitons or lumps, and being of the dark or of the anti-dark type. 
A brief description of our findings, as well as the outline of 
the presentation, is as follows. 
In Section~II, we present the model, and study both  the linear and the nonlinear regime. 
We present results of multiscale expansion methods that are used for the derivation  
of asymptotic reductions of the CH-NLS equation. In particular, at an intermediate 
stage of the asymptotic analysis, we obtain a 2D Boussinesq-type equation, and 
also obtain its far-field, namely a pair of KP equations (that are either of the 
KP-I or KP-II type) for right- and left-going waves. In Section~III, we use solutions 
of the KP-I and KP-II models to construct approximate soliton solutions of the 
original CH-NLS equation; the derived solutions have the form of dark or anti-dark 
line solitons and lumps. We also present results of direct numerical simulations 
concerning the dynamics and interactions between the various approximate soliton 
solutions. We find that, for sufficiently small amplitudes, all the derived 
solutions persist and can undergo quasi-elastic head-on collisions.
Finally, in Section~4, we summarize our findings and present our conclusions,
as well as suggest a number of directions for future study.

\section{Model and its analytical consideration}

\subsection{Dispersion relation, continuous-wave and its stability}

As indicated above, our aim is to study the 2D CH-NLS equation, which is a 
generalization of the 1D model derived in Refs.~\cite{Arnaudon,Arnaudon2}, 
when developing a theory of a deformation of hierarchies of integrable systems. 
The 2D CH-NLS equation is expressed as follows:
\begin{equation}\label{1}
im_t+\Delta u+2\sigma m(|u|^2-a^2|\boldsymbol{\nabla} u|^2)=0, \quad m=u-a^2\Delta u,
	\end{equation}
where $u(x,y,t)$ and $m(x,y,t)$ are complex fields,  
$\Delta=\partial_{x}^2+\partial_{y}^2$ is the Laplacian in 2D, and 
$\boldsymbol{\nabla}=(\partial_x,\partial_y)$ is the gradient operator. In addition, 
$\sigma= \pm 1$ pertains, respectively, to focusing or defocusing nonlinearity, 
while the constant $a$ arises within the Helmholtz operator; note that for $a=0$ 
the above model reduces to the standard, 2D NLS equation. I.e., this
parameter measures the ``size'' of the departure from the original
NLS limit.

In terms of the complex field $u$, the CH-NLS equation can be expressed as:
	\begin{equation}\label{2}
	iu_t+\Delta u+2\sigma u|u|^2-ia^2\Delta u_t-2\sigma a^2u|\boldsymbol{\nabla} u|^2
	-2\sigma a^2\Delta u|u|^2+2\sigma a^4\Delta u|\boldsymbol{\nabla} u|^2=0. \quad
	\end{equation}

The simplest nontrivial solution of Eq.~(\ref{2}) is the continuous-wave (cw): 
\begin{equation}
u=u_{0}\exp(2i\sigma|u_{0}|^{2}t), 
\label{cw}
\end{equation}
where $u_{0}$ is an arbitrary complex constant. Since below we will seek 
nonlinear excitations (e.g., solitary waves) which propagate on top of this
cw background, it is relevant to investigate if this solution is subject to modulational
instability (MI): naturally, nonlinear excitations corresponding to an unstable background will be less physically relevant. The stability of the cw solution (\ref{cw}) can 
be investigated upon using the ansatz $u=u_{0}(1+\tilde{u})\exp(2i\sigma|u_{0}|^{2}t+i\theta)$, 
where the small perturbations $\tilde{u}$ and $\theta$ 
are assumed to be $\propto \exp(i\boldsymbol{k} \cdot \boldsymbol{r}-i\omega t)$, 
with $\boldsymbol{r}=(x,y)$, while $\boldsymbol{k}=(k_{x},k_{y})$ and $\omega$ denote 
the perturbation wave-vector and frequency, respectively. Then, it can readily be found 
that $\omega$ and $k\equiv|\boldsymbol{k}|$ obey the following the dispersion relation:
	\begin{equation}\label{9}
	\omega^2=\frac{k^{2} (-4\sigma |u_0|^2+k^{2})}{(1+a^2k^{2})^2}.
	\end{equation}
It is observed that in the case of the defocusing nonlinearity, i.e., for $\sigma=-1$, 
the cw solution is always modulationally stable, i.e., 
$\omega \in \mathbb{R}~\forall k\in \mathbb{R}$. On the other hand, for a 
focusing nonlinearity, $\sigma=+1$, the cw solution is unstable for $k^2<4|u_0|^2$: 
in this case, perturbations grow exponentially, with the 
instability growth rate given by Im$(k)$. Note that, for $a = 0$, and in the 1D case 
(e.g., $k_{y}=0$), Eq.~(\ref{9}) reduces to the well-known~\cite{kivshar} 
result for the modulational (in)stability of the NLS equation: 
$\omega^{2}=k^{2}(-4\sigma|u_{0}|^{2}+k^{2})$.	
Clearly (and as was also found in the 1D case~\cite{Arnaudon2,2017}), 
the MI band is shared between NLS and CH-NLS and, in both cases, 
the cw~(\ref{cw}) is modulationally stable in the defocusing realm of $\sigma=-1$.	
For this case, it is also relevant to mention that the long-wavelength limit 
($k \rightarrow 0$) of Eq.~(\ref{9}) provides the (squared) ``sound velocity'', 
\begin{equation}
C^{2}= 4|u_{0}|^{2},
\label{C}
\end{equation}
namely the velocity of small-amplitude linear excitations propagating on top of the cw background.

\subsection{Multiscale expansions and reduced models}

We now consider small-amplitude slowly-varying modulations of the cw solution, 
and look for solutions of Eq.~(\ref{2}) in the form of the following asymptotic 
expansions:
\begin{eqnarray}
u&=&u_0\rho\exp\left[-2i|u_0|^2 t + i \epsilon^{1/2}\Phi(X,Y,T)\right], 
\label{u} \\
\rho&=& 1+\sum_{j=1}^{\infty}\epsilon^j \rho_j(X,Y,T), 
\label{rho} 
\end{eqnarray}
where the phase $\Phi$ and densities $\rho_j$ are unknown real functions of the slow variables
\begin{equation}
X=\epsilon^{1/2}x, \quad Y=\epsilon^{1/2}y, \quad T=\epsilon^{1/2}t,
\end{equation}
%
while $0<\epsilon \ll1$ is a formal small parameter. 
Substituting the expansions~(\ref{u})-(\ref{rho}) into Eq.~(\ref{2}), and separating real 
and imaginary parts, we obtain the following results. 
First, the real part of Eq.~(\ref{2}) leads, 
at orders $\mathcal{O}(\epsilon)$ and $\mathcal{O}(\epsilon^{2})$, 
to the following equations:
	\begin{eqnarray}\label{13}
	\Phi_{T} +C^2 \rho_{1}=0,
	\end{eqnarray}
	\begin{eqnarray}\label{14}
	a^{2} \tilde{\Delta} \Phi_{T} -2|u_{0}|^{2}(2\rho_{2}+3\rho_{1}^{2}
	-a^{2}|\boldsymbol{\tilde{\nabla}} \Phi|^{2}) 
	+\tilde{\Delta} \rho_{1} -\rho_{1}\Phi_{T}
	-|\boldsymbol{\tilde{\nabla}}\Phi|^{2}=0,
	\end{eqnarray}
where $\tilde{\Delta} \equiv \partial_X^2 + \partial_Y^2$ and   
$\tilde{\boldsymbol{\nabla}}\equiv (\partial_X,\partial_Y)$.	
Second, the imaginary part of Eq.~(\ref{2}), at orders $\mathcal{O}(\epsilon^{3/2})$ 
and $\mathcal{O}(\epsilon^{5/2})$ yields:
	\begin{eqnarray}\label{15}
	\rho_{1T}+\tilde{\Delta} \Phi=0,
	\end{eqnarray}
	\begin{eqnarray}\label{16}
	\rho_{2T}
	%
	+(1+a^{2}C^{2})\rho_{1}\tilde{\Delta}\Phi
	+a^{2}\Phi_{T}\tilde{\Delta}\Phi-a^{2}\tilde{\Delta} \rho_{1T} 
	+2\tilde{\nabla}\Phi \cdot ( a^2\tilde{\boldsymbol{\nabla}}\Phi_{T} +\tilde{\nabla}\rho_{1})
	=0.
	\end{eqnarray}
Then, eliminating the functions $\rho_{1}$ and $\rho_2$ from the system 
of Eqs.~(\ref{13})-(\ref{16}), we derive the following equation for $\Phi(X,Y,T)$:
	\begin{eqnarray}\label{17}
	\Phi_{TT}-C^2 \tilde{\Delta} \Phi+\epsilon \Big\{2a^{2}\tilde{\Delta}\Phi_{TT}
	 -4(1-3a^{2}|u_{0}|^{2})( \tilde{\boldsymbol{\nabla}}  \Phi_{T} \cdot \tilde{\boldsymbol{\nabla}}\Phi)
	 -2\Phi_{T}\tilde{\Delta}\Phi
	-\tilde{\Delta}^2 \Phi) 
	\Big \}= \mathcal{O}(\epsilon^{2}).
	\end{eqnarray}
At the leading-order, Eq.~(\ref{17}) is a second-order linear wave 
equation, while at order $\mathcal{O}(\epsilon)$ it incorporates fourth-order
dispersion and quadratic nonlinear terms, similar to the 
Boussinesq and the Benney-Luke \cite{BL} equations. These models 
have originally been used to describe bidirectional shallow 
water waves~\cite{ablo2011}, but also ion-acoustic 
waves in plasmas \cite{infeld}, as well as 
mechanical lattices and electrical transmission lines \cite{rem}. 
Note that the present analysis generalizes the results of Ref.~\cite{2017} (where the 1D case was 
studied) to the 2D setting; in that regard, it is worth mentioning that similar 
2D Boussinesq equations were derived from 2D NLS equations with either 
a local \cite{pel1,pel2} or a nonlocal \cite{r6,2d3} defocusing nonlinearity. 

Next, using a multiscale expansion method similar to the one employed in the water 
wave problem \cite{ablo2011}, we will derive the far-field of the Boussinesq equation, 
namely a pair of two KP equations for right- and left-going waves. These models 
will be obtained under the additional assumption of unidirectional propagation. 
We thus seek solutions of Eq.~(\ref{17}) in the form of the asymptotic expansion:
\begin{equation}\label{19}
\Phi=\Phi_{0}+\epsilon\Phi_{1}+\cdots, 
\end{equation}
where the unknown functions $\Phi_j$ ($j=1,2,\ldots$) depend on the variables:
\begin{equation}\label{18}
\mathcal{\chi}=X-CT, \qquad \tilde{\mathcal{\chi}}=X+CT, \qquad \mathcal{Y}=\sqrt{\epsilon}Y, \qquad \mathcal{T}=\epsilon T , 
\end{equation}
%
Substituting Eq.~(\ref{19}) into Eq.~(\ref{17}), 
and using Eq.~(\ref{18}), we obtain the following results. 
First, at the leading order, $\mathcal{O}(1)$, we obtain the wave equation:
	\begin{equation}\label{20}
	-4C^{2}\Phi_{0 \mathcal{\chi} \tilde{\mathcal{\chi}}}=0, 
	\end{equation}
which implies that $\Phi_{0}$ can be expressed as a superposition of a right-going wave, 
$\Phi_{0}^{ (R) }$, depending on $\mathcal{ \chi }$, and a left-going wave, $\Phi_{0}^{(L)}$, 
depending on $\tilde{ \mathcal{\chi} }$, namely:
	\begin{equation}\label{21}
	\Phi_{0}=\Phi_{0}^{(R)}+\Phi_{0}^{(L)}.
	\end{equation}
Second, at $\mathcal{O}(\epsilon)$, we obtain the equation:
	\begin{eqnarray}\label{22}
	4C^{2}\Phi_{1 \mathcal{\chi} \tilde{\mathcal{\chi}}} = C(3a^{2}C^{2}-2)(\Phi_{0 \mathcal{\chi}}^{(R)}\Phi_{0 \tilde{\mathcal{\chi}} \tilde{\mathcal{\chi}}}^{(L)}-
	\Phi_{0 \tilde{\mathcal{\chi}}}^{(L)}\Phi_{0 \mathcal{\chi} \mathcal{\chi}}^{(R)})
	\nonumber\\
	+\Bigg \{ \Big[2C\Phi_{0 \mathcal{T}}^{(L)} -\frac{3C}{2}(2-a^{2}C^{2})\Phi_{0 \tilde{\mathcal{\chi}}}^{(L)2} -(1-2a^{2}C^{2})\Phi_{0 \tilde{\mathcal{\chi}} \tilde{\mathcal{\chi}} \tilde{\mathcal{\chi}}}^{(L)}  \Big]_{\tilde{\mathcal{\chi}}} -C^{2}\Phi^{(L)}_{0 \mathcal{Y} \mathcal{Y}} \Bigg \}
	\nonumber\\
	-\Bigg \{ \Big[2C\Phi_{0 \mathcal{T}}^{(R)} -\frac{3C}{2}(2-a^{2}C^{2})\Phi_{0 \mathcal{\chi}}^{(R)2} +(1-2a^2 C^{2})\Phi_{0 \mathcal{\chi} \mathcal{\chi} \mathcal{\chi}}^{(R)}\Big]_{\mathcal{\chi}} +C^{2}\Phi^{(R)}_{0 \mathcal{Y} \mathcal{Y}} \Bigg \}. 
	\end{eqnarray}
When integrating Eq.~(\ref{22}) with respect to $\mathcal{\chi}$ or 
$\tilde{\mathcal{\chi}}$, it is observed that secular terms arise from the curly brackets 
in the right-hand side of this equation, which are functions of $\mathcal{\chi}$ or 
$\tilde{\mathcal{\chi}}$ alone, not both. Hence, these secular terms are set to zero, which leads 
to two uncoupled nonlinear evolution equations for $\Phi_{0}^{(R)}$ and $\Phi_{0}^{(L)}$. 
Next, employing Eq.~(\ref{13}), it is found that the amplitude $\rho_{1}$ can also be 
decomposed to a left- and a right-going wave, i.e., $\rho_{1}=\rho_{1}^{(R)}+\rho_{1}^{(L)}$, 
with
	\begin{equation}\label{23}
	\Phi_{0 \mathcal{\chi}}^{(R)}=C\rho_{1}^{(R)}, \quad
	\Phi_{0 \tilde{\mathcal{\chi}}}^{(L)}=-C\rho_{1}^{(L)}.
	\end{equation}
        Using these and
        the above equations for $\Phi_{0}^{(R)}$ and $\Phi_{0}^{(L)}$ yields 
the following equations for $\rho_{1}^{(R)}$ and $\rho_{1}^{(L)}$:
	\begin{eqnarray}\label{24}
	\Big [ 2C\rho_{1 \mathcal{T}}^{(R)} -3C^{2}(2-a^{2}C^{2})\rho_{1}^{(R)} \rho_{1 \mathcal{\chi}}^{(R)}+(1-2a^{2}C^{2})\rho_{1 \mathcal{\chi} \mathcal{\chi} \mathcal{\chi} }^{(R)} \Big ]_{\mathcal{\chi}} +C^{2} \rho_{1 \mathcal{Y} \mathcal{Y}}^{(R)}  =0,
	\end{eqnarray}
	\begin{eqnarray}\label{25}
 \Big [  2C\rho_{1 \mathcal{T}}^{(L)} +3C^{2}(2-a^{2}C^{2})\rho_{1}^{(L)} \rho_{1 \tilde{\mathcal{\chi}}}^{(L)} -(1-2a^{2}C^{2})\rho_{1 \tilde{\mathcal{\chi}} \tilde{\mathcal{\chi}} \tilde{\mathcal{\chi}} }^{(L)} \Big ]_{\tilde{\mathcal{\chi}}} -C^{2} \rho_{1 \mathcal{Y} \mathcal{Y}}^{(L)} =0.
	\end{eqnarray}
        The result of this analysis is
        the emergence of two KP equations for left- and right-going waves. 
Without loss of generality, below we focus on the right-going wave. 
To proceed further, and express the KP of Eq.~(\ref{24}) in its standard form, we 
introduce the rescaling: 
	\begin{eqnarray}\label{26}
	\mathcal{\hat{Y}} = \sqrt{\frac{3 |1-2a^{2}C^{2}|}{C^{2}}}\mathcal{Y}, \qquad \mathcal{\hat{T}}=\Big(\frac{1-2 a^{2}C^{2}}{2C}\Big)\mathcal{T}, \qquad 
	\rho_{1}^{(R)}= qU,
	\end{eqnarray}
where the parameter $q$ is given by:
	\begin{equation}\label{31}
	q=\frac{2}{C^{2}}\left(\frac{1-2 a^{2}C^{2}}{a^{2}C^{2}-2}\right).
	\end{equation}
Then, Eq.~(\ref{24}) is reduced to the form:
	\begin{equation}\label{27}
	\Big ( U_{\mathcal{\hat{T}}} +6UU_{\mathcal{\chi}} +U_{\mathcal{\chi} \mathcal{\chi} \mathcal{\chi}} \Big)_{\mathcal{\chi}} +3 \delta U_{\mathcal{\hat{Y}} \mathcal{\hat{Y}}}=0,
	\end{equation}
where $\delta=- {\rm sgn}(1-2a^2C^2) = \pm 1$. The case $\delta=-1$, or  $1-2a^2C^2>0$, 
corresponds to the KP-I equation, which models waves in liquid thin films with large surface tension. 
On the other hand, $\delta=1$, or $1-2a^2C^2<0$, corresponds to 
the KP-II equation, arising in the description of shallow water waves characterized 
by small surface tension (see, e.g., \cite{ablo2011}). 
	
\section{Line solitons and lumps}

Below, soliton solutions of the KP-I and KP-II models will be used for the construction 
of approximate soliton solutions of the original 2D CH-NLS equation. Furthermore, the dynamics 
of these structures will be studied by means of direct numerical simulations in the framework 
of the CH-NLS. 
For this  numerical exploration presented below, we note the following. 
For the numerical integration of the original 2D CH-NLS,
we use the Exponential Time-Differencing 4th-order Runge-Kutta 
(ETDRK4) scheme of Refs.~\cite{7,8}; for details related to implementation, 
cf. Ref.~\cite{2017}. The parameters used in the simulations can be found in the individual figure captions. If a figure shows a collision between two solitons, and only one set of parameters is given, then that set was used for both solitons. Lastly, the background amplitude has been set to unity for all simulations.

\subsection{Approximate soliton solutions of the CH-NLS equation}
	
We start with the case of line soliton solutions, which are supported by both KP-I and KP-II 
equations, given their quasi-1D nature, [cf. Eq.~(\ref{27})] and are of the form~\cite{ablo91,ablo2011}:
	\begin{equation}\label{28}
U=\frac{1}{2} \gamma^{2} \; {\rm sech}^{2} \Big[\frac{\gamma}{2}(\mathcal{\chi}-\beta\mathcal{\hat{Y}} -\frac{\eta}{\gamma}\mathcal{\hat{T}}+\mathcal{\chi}_{0}) \Big],
	\end{equation} 
where $\eta,\gamma$ and $\beta$ are constants with $\eta=\gamma^{2}+3\delta\gamma\beta^{2}$, 
and $\delta=\pm 1$ for KP-I or KP-II, respectively. The solution~(\ref{28}) is characterized by the parameter $\gamma$ associated with
the soliton amplitude and the soliton direction $\beta$ in the $xy$-plane, with $\beta=\tan(\gamma)$. Using Eq.~(\ref{28}), and reverting transformations back to the original 
variables, we find the following approximate (valid up to order $O(\epsilon)$) solution of 
the CH-NLS, Eq.~(\ref{1}):
	\begin{eqnarray}\label{29}
	u \approx u_{0}\Big[1+ \frac{\epsilon\gamma^{2} q}{2} ~{\rm sech}^2(\xi) \Big] 
	\exp\Big[-2i|u_{0}|^{2} t 
	+ i \sqrt{\epsilon} C \gamma q \tanh(\xi) \Big], 
	\end{eqnarray}
where
	\begin{equation}\label{30}
	\xi=\frac{\sqrt{\epsilon} \gamma}{2} 
	\left[ x-\sqrt{\epsilon}\beta \sqrt{\frac{3|1-2a^{2}C^{2}|}{C^{2}}}y-\left(C+\frac{\epsilon\eta(1-2a^{2} C^{2})}{2\gamma C} \right)t+x_{0} \right].
	\end{equation}
Here, it is important to mention that the approximate soliton solution~(\ref{29}) 
describes two types of solitons: if $q>0$ the solitons are dark, having the form of  
density dips on top of the cw background of amplitude $u_0$; if $q<0$, the 
solitons are anti-dark, having the form of density humps on top of the cw background. 
The sign of parameter $q$ 
depends on the range of values of a single parameter $p \equiv a^2 C^2 = 4a^2|u_0|^2$: 
indeed, if $1/2 < p <2$ the solitons are antidark, else they are dark. 

	\begin{figure}[tbp]
		\subfloat[t=0]{\includegraphics[width=.26\textwidth]{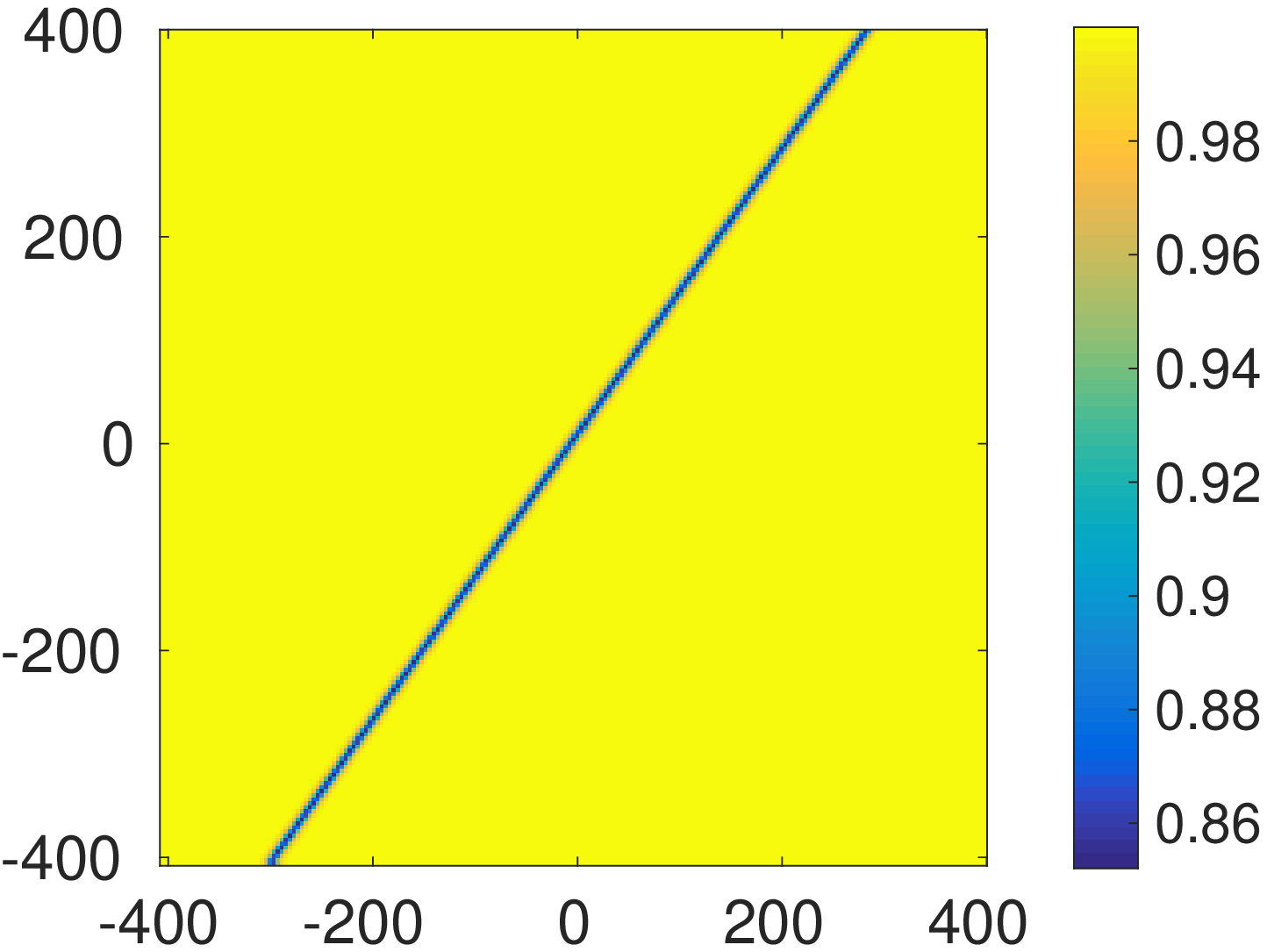}}\quad
		\subfloat[t=100]{\includegraphics[width=.26\textwidth]{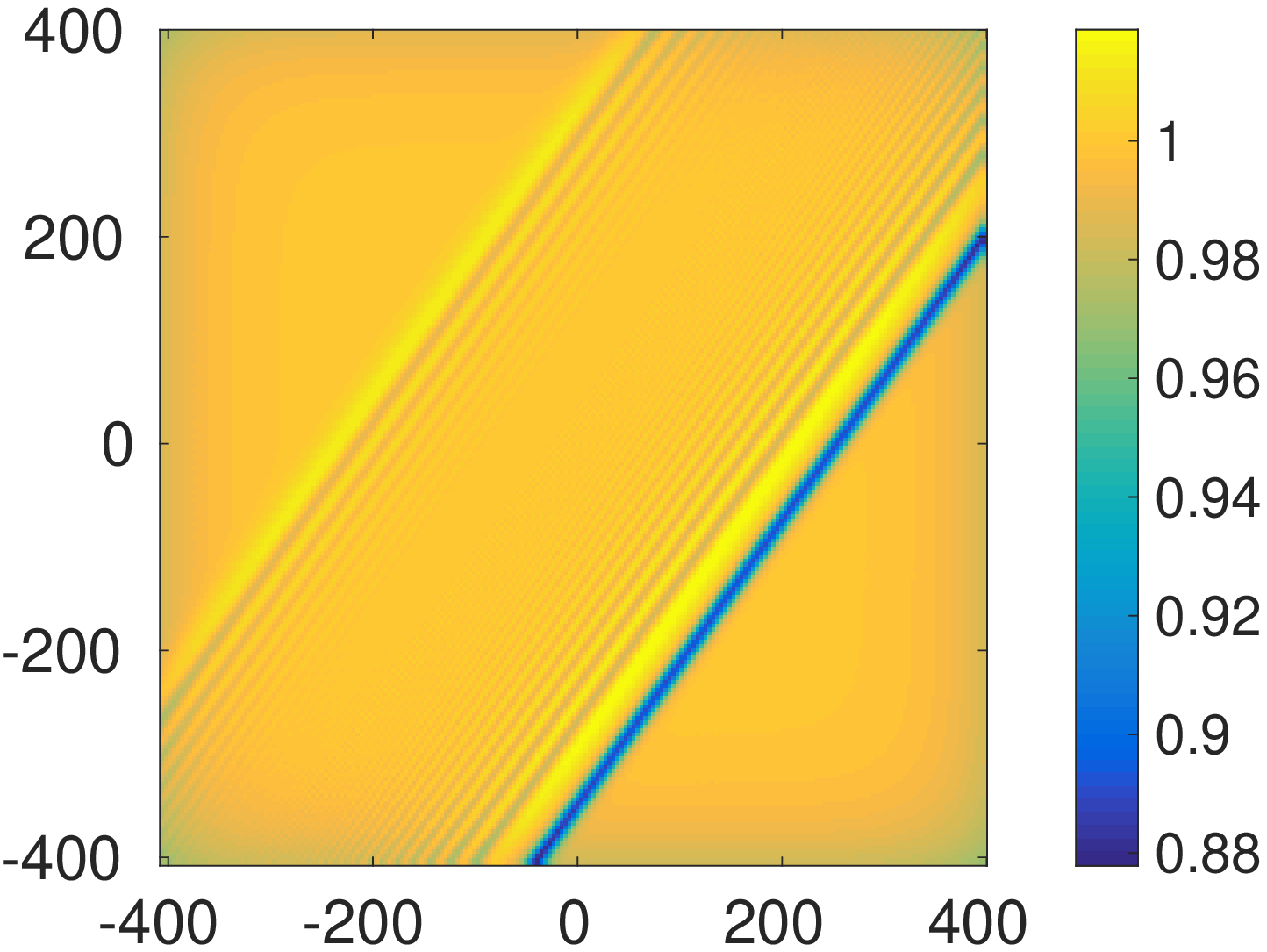}} \\
		\subfloat[t=0]{\includegraphics[width=.26\textwidth]{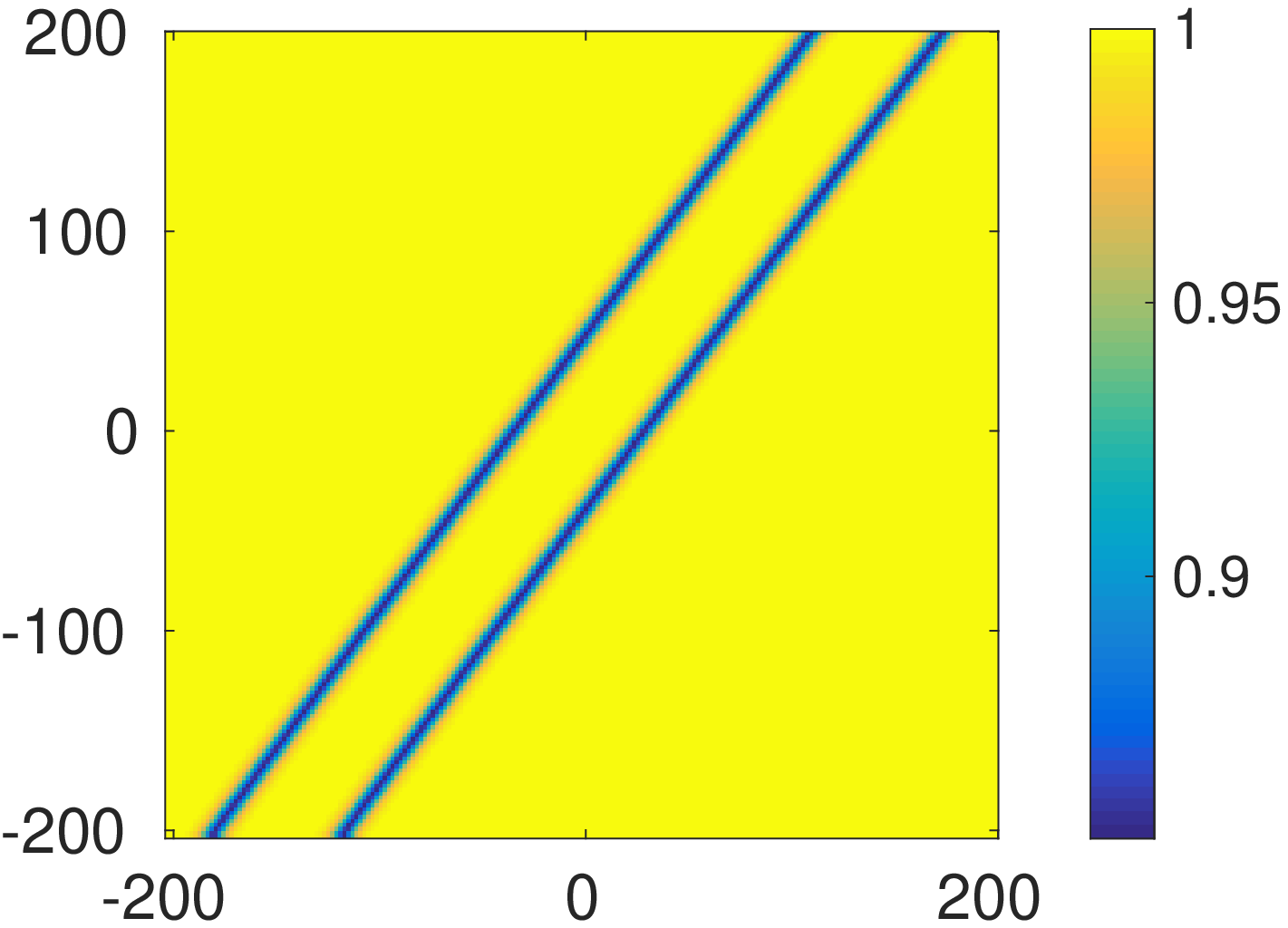}}\quad
	\subfloat[t=100]{\includegraphics[width=.26\textwidth]{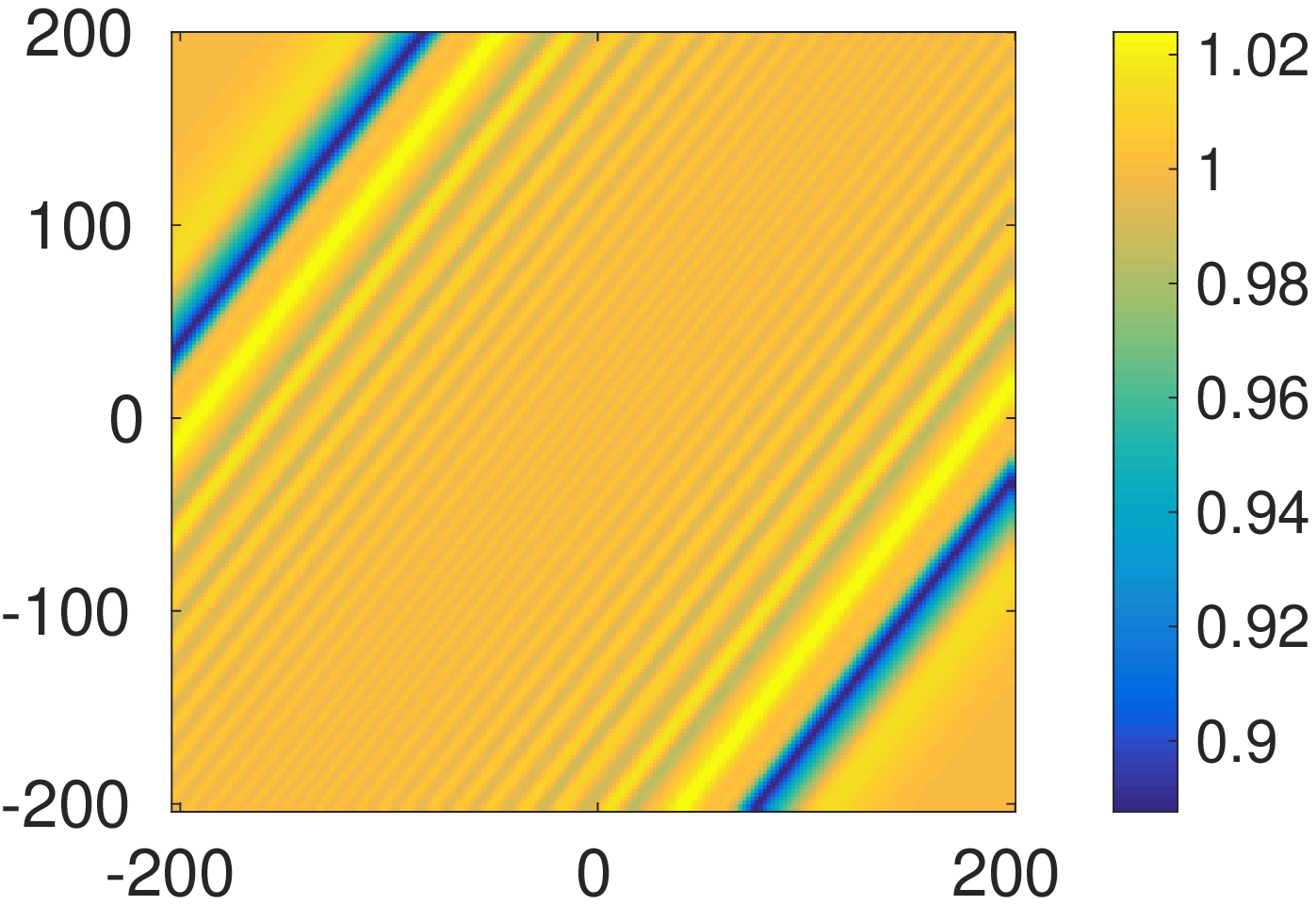}}
\caption{(Color online) 
Top panels: Contour plots showing the evolution of the density of one line dark soliton 
of relatively large amplitude, with $\epsilon=0.1$, at $t=0$ [panel (a)] and $t=100$ [panel (b)]. 
It is observed that the soliton splits into two waves, a dark and an anti-dark, and also emits radiation. 
Bottom panels: Collision between two line dark solitons. Panel (c) shows the initial condition, at $t=0$, 
and panel (d) shows the outcome of the head-on collision, at $t=100$. Here, the leftmost soliton 
appears as the rightmost one, and vice versa. 
Parameter values: $\alpha=1$, $\gamma=1.3$, and $\beta=1$.} \label{Fig4}
	\end{figure}

Using direct numerical simulations (results not shown here) we have found that for sufficiently 
small $\epsilon$, of the order $\mathcal{O}(10^{-2})$, solitons of both types do exist and 
propagate undistorted, without emitting significant radiation.  
%
%
%
On the other hand, solitons of relatively large amplitudes feature a different behavior, 
because -- as expected -- the results of the asymptotic analysis become less accurate. Indeed, 
this is shown in Fig.~\ref{Fig4}, where the evolution of a dark line soliton 
[cf. initial condition, at $t=0$, in panel (a)] is depicted. As 
is observed in panel~(b), the dark soliton ``ejects'' an anti-dark line soliton, 
and a radiation tail forms. Nevertheless, it should be pointed out that 
even for such relatively large amplitudes, the approximate soliton solutions are supported by 
the system and can even undergo almost elastic collisions with each other.
An example 
pertaining to the case of a pair of dark line solitons is shown in panels~(c) and (d) of Fig.~\ref{Fig4}. 
Here, the collision is deemed as almost elastic, in the sense that 
the observed dynamics for each soliton is identical to the one depicted 
in panel (b), i.e., the ejection of the anti-dark soliton and the emission of radiation would occur 
even if each of the solitons evolved by itself (i.e.,
in the absence of the other one).



	\subsection{Lump solitons}

	\begin{figure}[tbp]
		\subfloat[t=0]{\includegraphics[width=.26\textwidth]{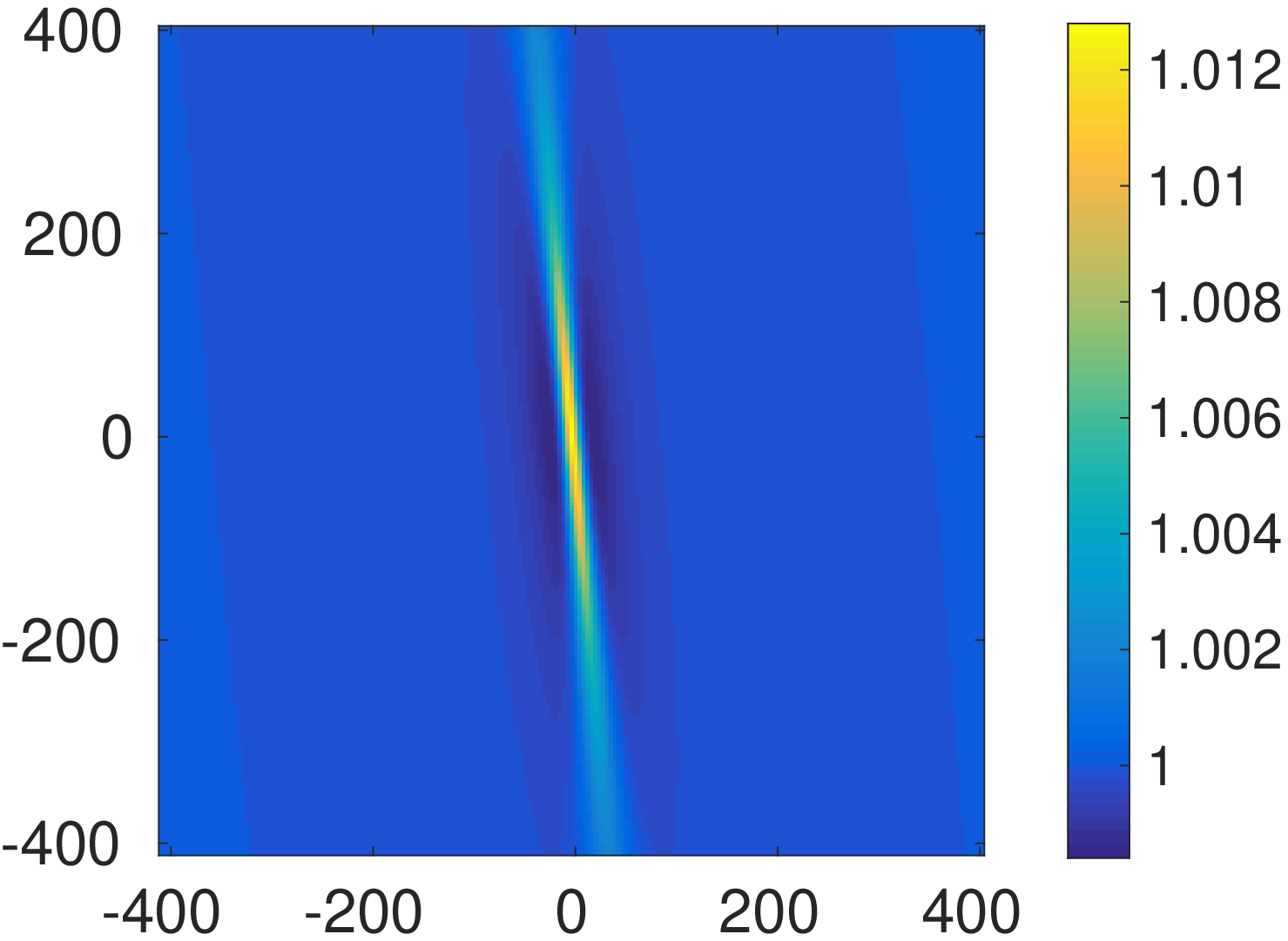}}\quad
		\subfloat[t=100]{\includegraphics[width=.26\textwidth]{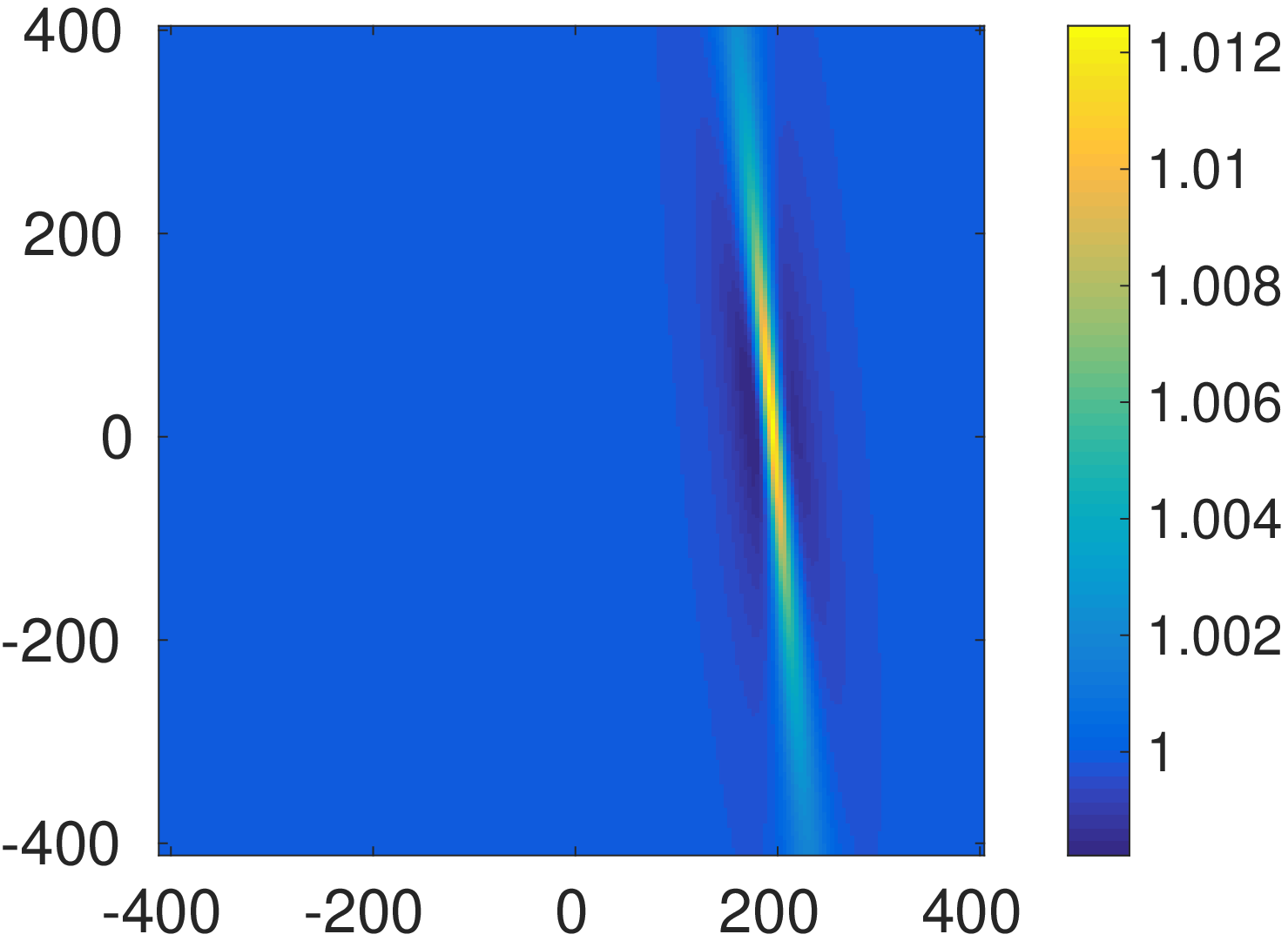}} \\
		\subfloat[t=0]{\includegraphics[width=.26\textwidth]{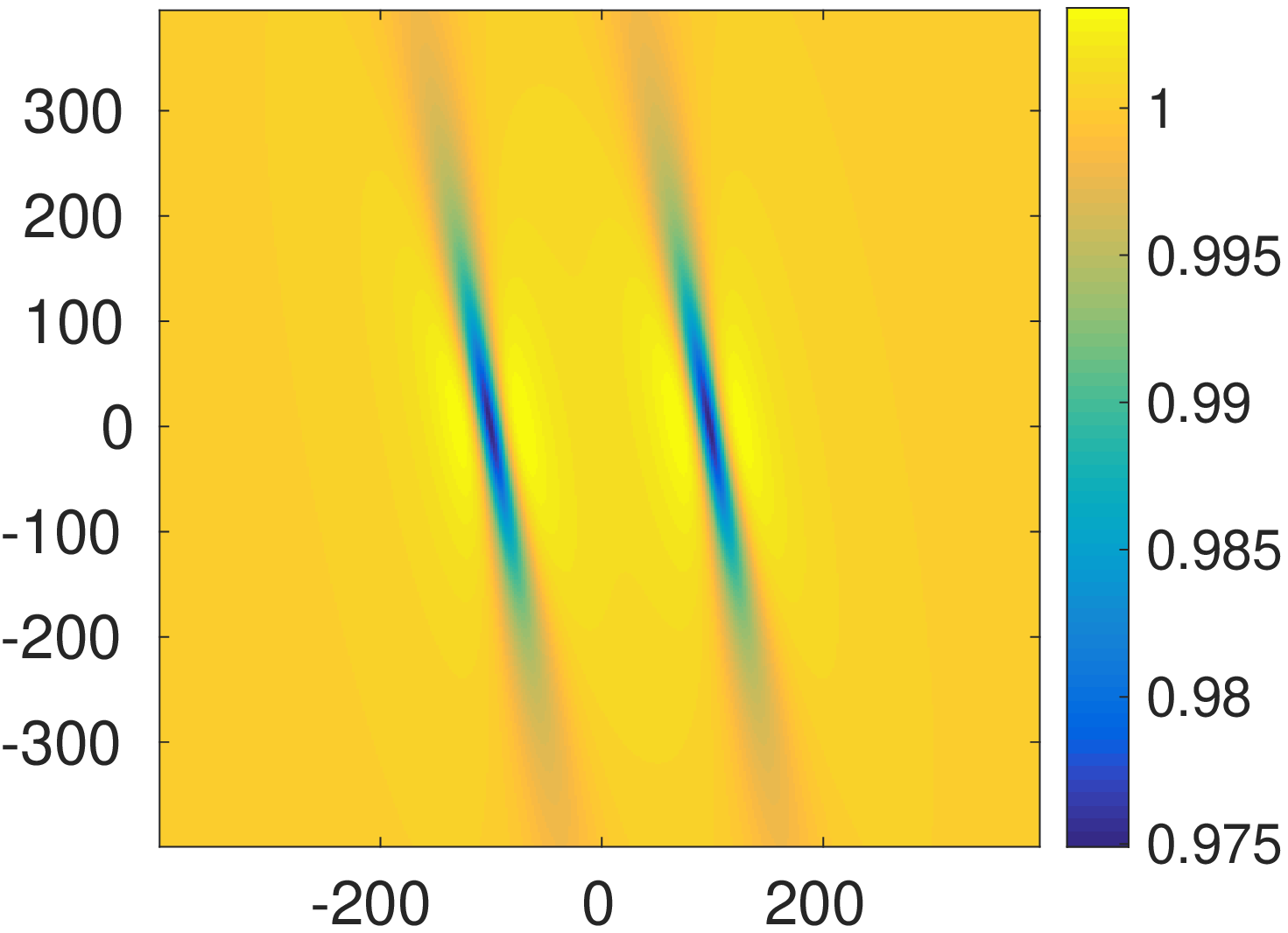}}\quad
		\subfloat[t=100]{\includegraphics[width=.26\textwidth]{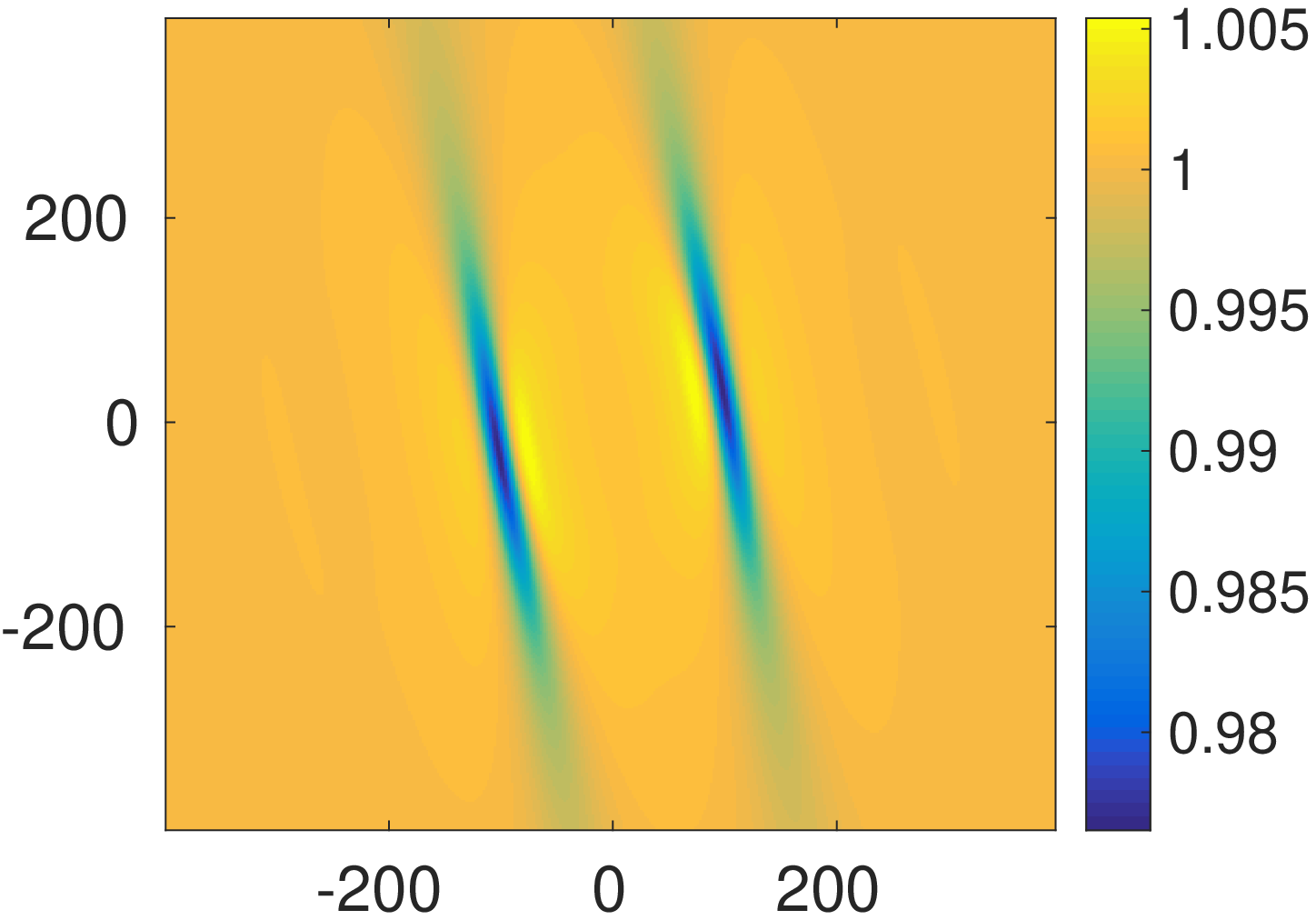}}
		\caption{(Color online) 
Top panels: Contour plots showing the evolution of the density of one anti-dark lump soliton 
of relatively small amplitude, with $\epsilon=0.01$, at $t=0$ [panel (a)] and $t=100$ [panel (b)]. 
It is observed that the lump evolves undistorted and no emission of radiation is observed. 
Parameter values: $\alpha=\frac{1}{2}$, $\gamma=1$, $\beta=0.8$.
Bottom panels: Collision between two dark lumps. Panel (c) shows the initial condition, at $t=0$, 
and panel (d) shows the outcome of the head-on collision, at $t=100$. Here, the leftmost lump 
appears as the rightmost one, and vice versa; again, the radiation
is barely discernible at $t=100$. 
Parameter values: $\alpha=\frac{1}{2}$, $\epsilon=0.005$, $\gamma=1$, $\beta=0.85$.
} \label{Fig2}
	\end{figure}

Next, we consider lump solitons, which are weakly localized (exponentially decaying) 
two-dimensional soliton solutions of the KP-I equation. The lump soliton is of the form:
	\begin{equation}\label{32}
		U= 4\frac{-\Big[ \mathcal{\chi}+\gamma \mathcal{\hat{Y}} +3(\gamma^{2}-\beta^{2}) \mathcal{\hat{T}} \Big ]^{2} +\beta^{2}(\mathcal{\hat{Y}}+6\gamma \mathcal{\hat{T}})^{2}+1/\beta^{2}}{ \Big \{ \Big[\mathcal{\chi}+\gamma \mathcal{\hat{Y}} +3(\gamma^{2}-\beta^{2}) \mathcal{\hat{T}} \Big ]^{2} +\beta^{2}(\mathcal{\hat{Y}}+6\gamma \mathcal{\hat{T}})^{2}+1/\beta^{2} \Big \}^{2} },
	\end{equation}
where $\gamma$ and $\beta$ are real parameters. As in the case of the line soliton, 
we use this solution and revert transformations back to the original variables, and 
find the following approximate (valid up to order $O(\epsilon)$) solution of Eq.~(\ref{1}):
\begin{eqnarray}\label{33}
u \approx u_{0}(1+ \epsilon q w_{1}) 
\exp[-2i|u_{0}|^{2} t 
+ i \sqrt{\epsilon} C q w_{2} ], 
\end{eqnarray}
where
\begin{eqnarray}\label{34}
w_{1}&=&4\frac{z_{-}+\frac{1}{\sqrt{\epsilon}\beta^{2}}}{\sqrt{\epsilon} \Big [  z_{+} +\frac{1}{\sqrt{\epsilon}\beta^{2}} \Big ]^{2}}, \nonumber \\
 w_{2}&=&4\frac{z_{0}}{\epsilon^{3/2} \beta^{2} \Big ( 6\gamma \sqrt{\epsilon} \frac{1-2a^{2}C^{2}}{2C} t +\sqrt{\frac{3|1-2a^{2}C^{2}|}{C^{2}}}y \Big )^{2} +z_{0}+\frac{1}{\sqrt{\epsilon}\beta^{2}}},
\end{eqnarray}
and
\begin{eqnarray}\label{35}
	z_{\pm}&=& \pm \Big [ x+\gamma\sqrt{\frac{3\epsilon|1-2a^{2}C^{2}|}{C^{2}}}y- \Big (C-\frac{3\epsilon(\gamma^{2}-\beta^{2})(1-2a^{2}C^{2})}{2C} \Big ) t \Big ]^{2}+\beta^{2} \Big[ \sqrt{\frac{3\epsilon|1-2a^{2}C^{2}|}{C^{2}}}y+\frac{6\epsilon \gamma(1-2a^{2}C^{2}) }{2C}t \Big]^{2}
\end{eqnarray}

In the case of small-amplitude lumps, e.g., for $\epsilon=0.01$, direct numerical simulations 
are in very good agreement with the analytical findings. Indeed, in Fig.~\ref{Fig2}, which 
depicts the evolution of an anti-dark lump, shown is the initial condition, at $t=0$ [panel (a)], 
and a snapshot, at $t=100$ [panel (b)], as found in the framework of the CH-NLS equation. 
It is observed that, up to this time, the lump soliton propagates undistorted and the radiation 
emitted is practically non observable. Furthermore, in the bottom panels of Fig.~\ref{Fig2}, shown is 
the result of a collision between two identical dark lumps -- one traveling to the 
left and one to the right. In panel (c), we depict the initial condition ($t=0$), 
and in panel (d) the outcome of the collision (at $t=100$), where the leftmost lump 
appears at the rightmost place, and vice versa. It is seen that, for such small-amplitudes, 
the two lump solitons remain unscathed after the collision while, in this case too, 
no radiation is observable. We note in passing that the validity of our analytical 
approximations was also checked in other cases (not shown here), e.g., for individual 
anti-dark line solitons and dark lump solitons, as well as for collisions between such 
structures, and -- for sufficiently small amplitudes -- a very good agreement with the 
numerical results was found as well.
	
%
%
%

\begin{figure}[tbp]
		\subfloat[t=0]{\includegraphics[width=.26\textwidth]{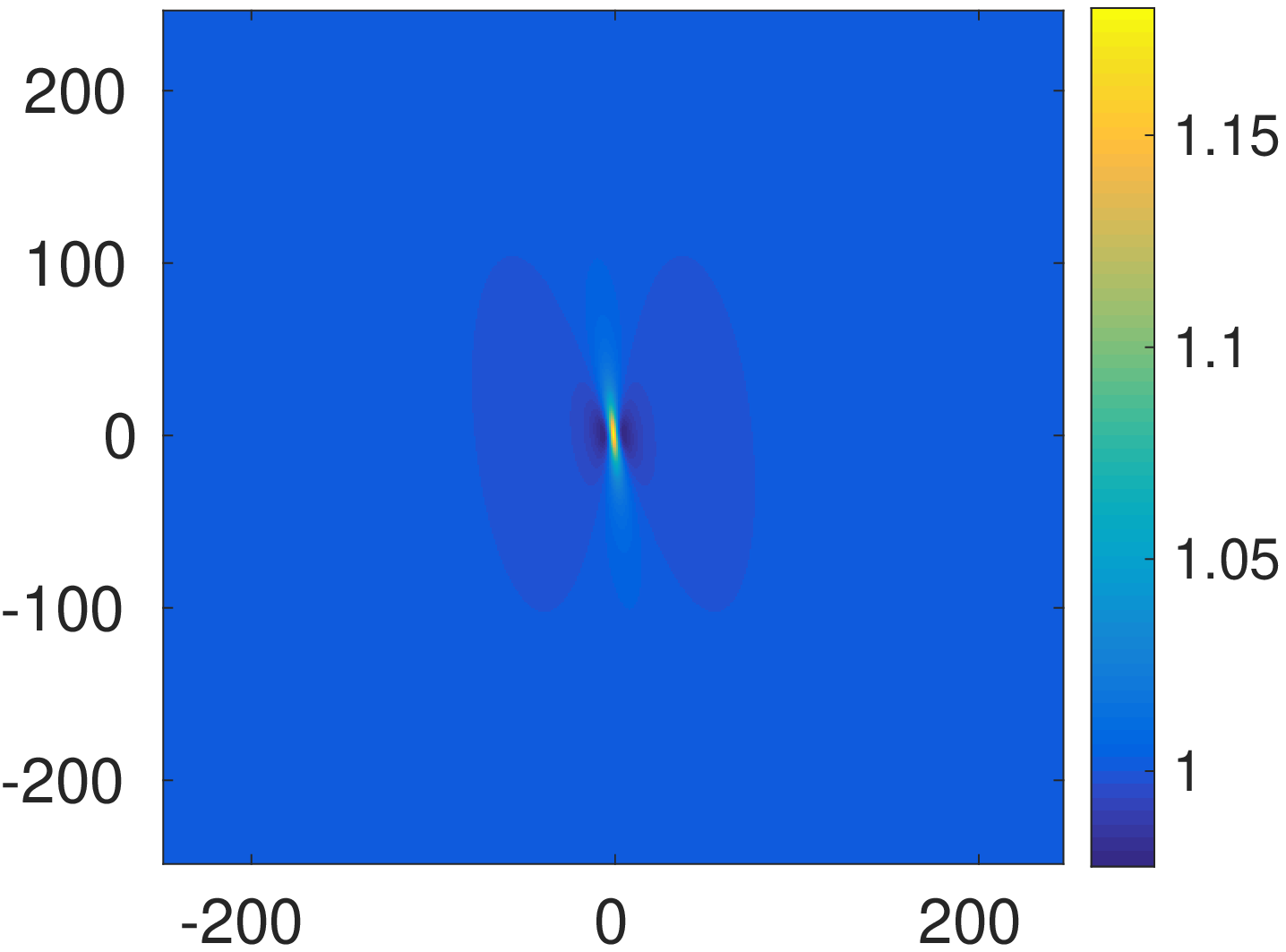}}\quad
		\subfloat[t=66]{\includegraphics[width=.26\textwidth]{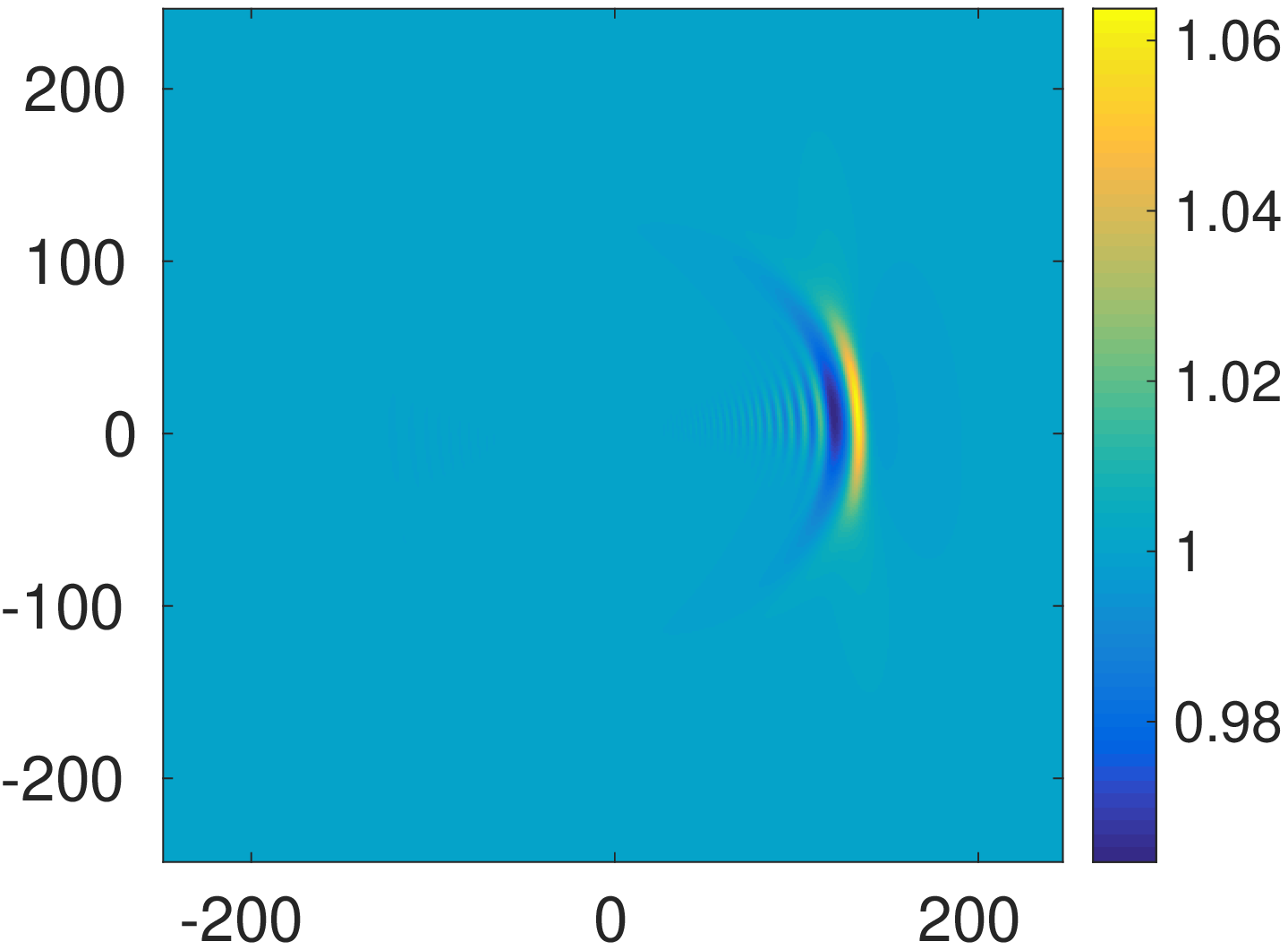}}\quad
		\subfloat[t=100]{\includegraphics[width=.26\textwidth]{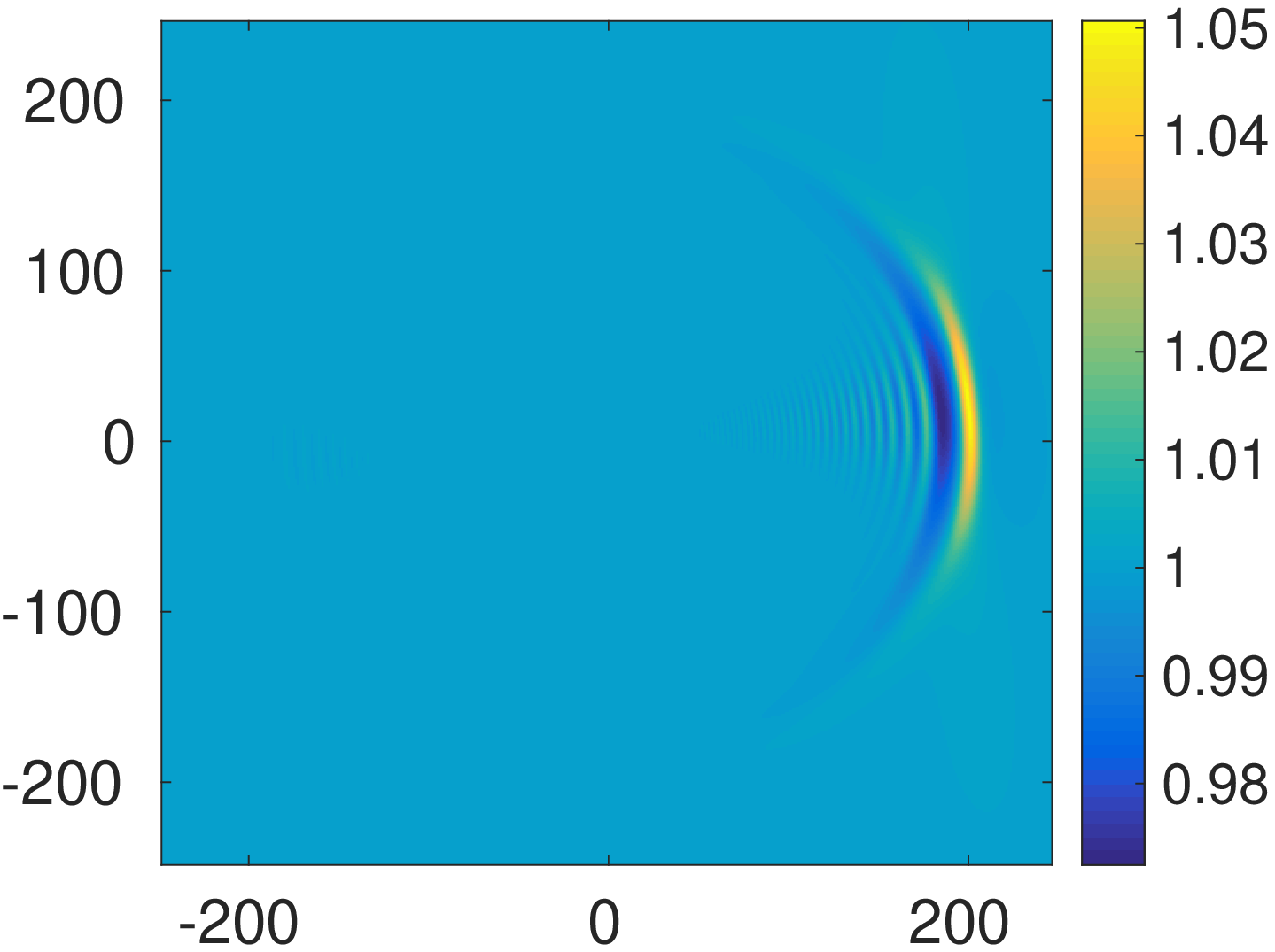}} \\
	\subfloat[t=0]{\includegraphics[width=.26\textwidth]{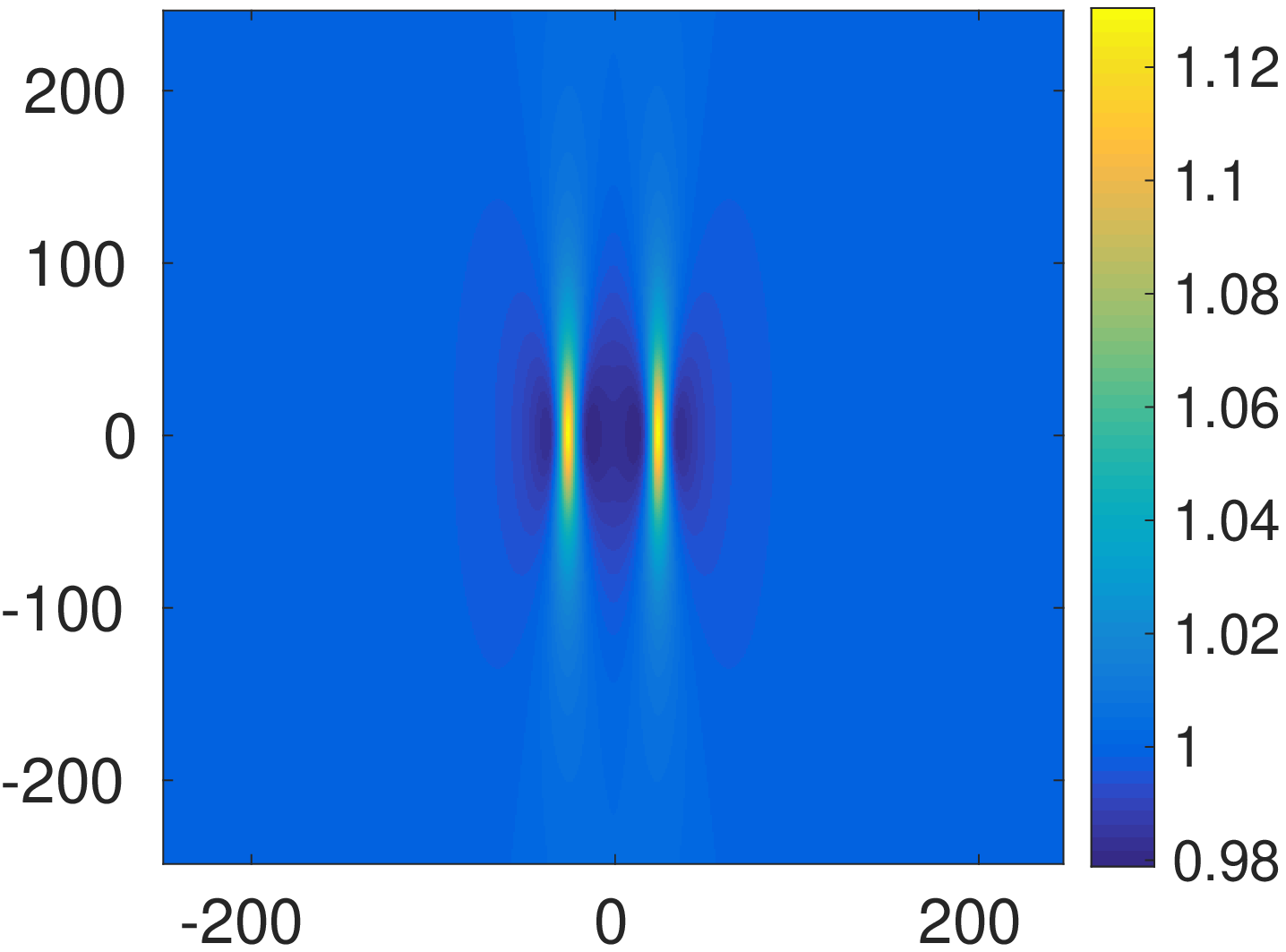}}\quad
	\subfloat[t=66]{\includegraphics[width=.26\textwidth]{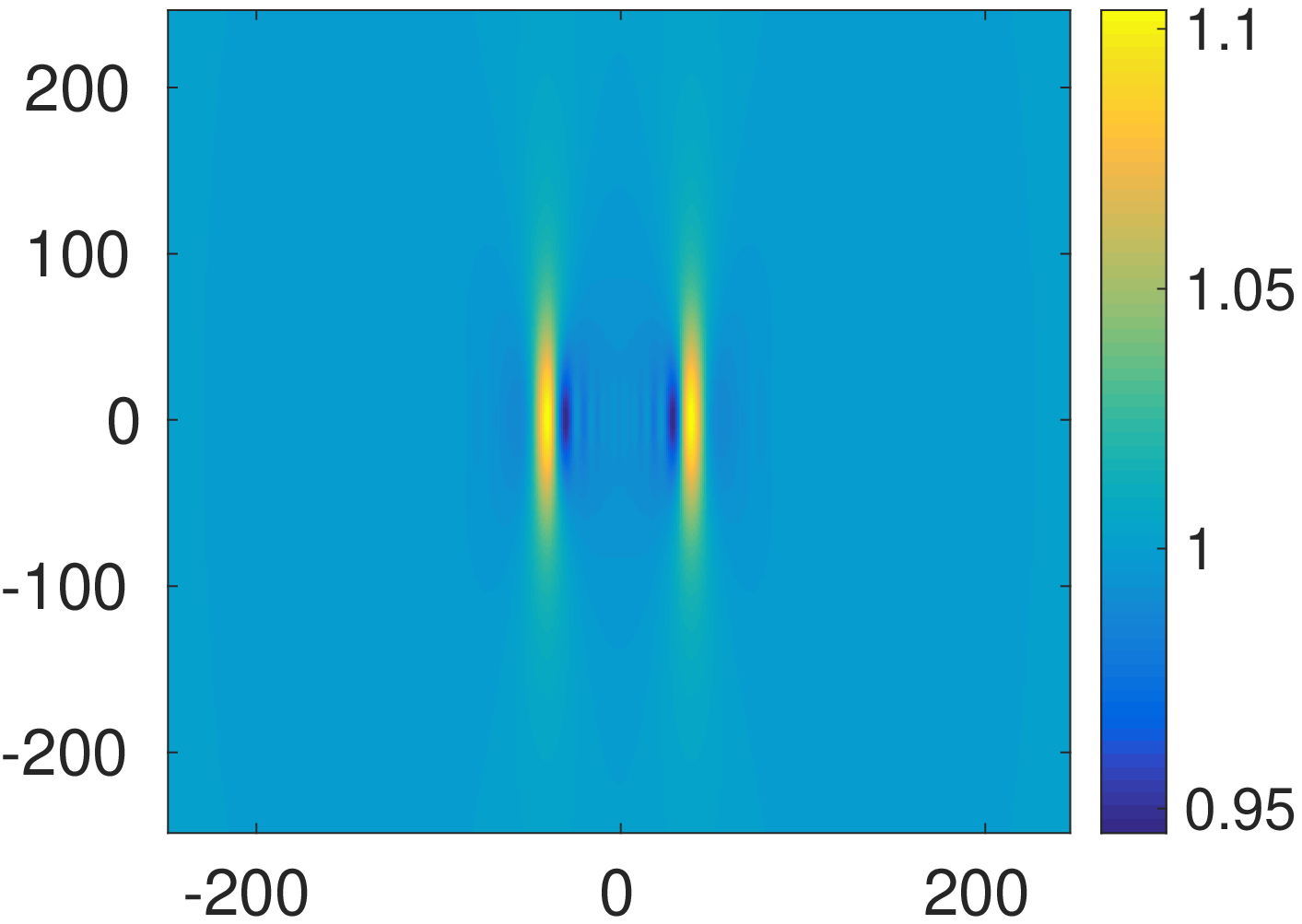}}\quad
	\subfloat[t=100]{\includegraphics[width=.26\textwidth]{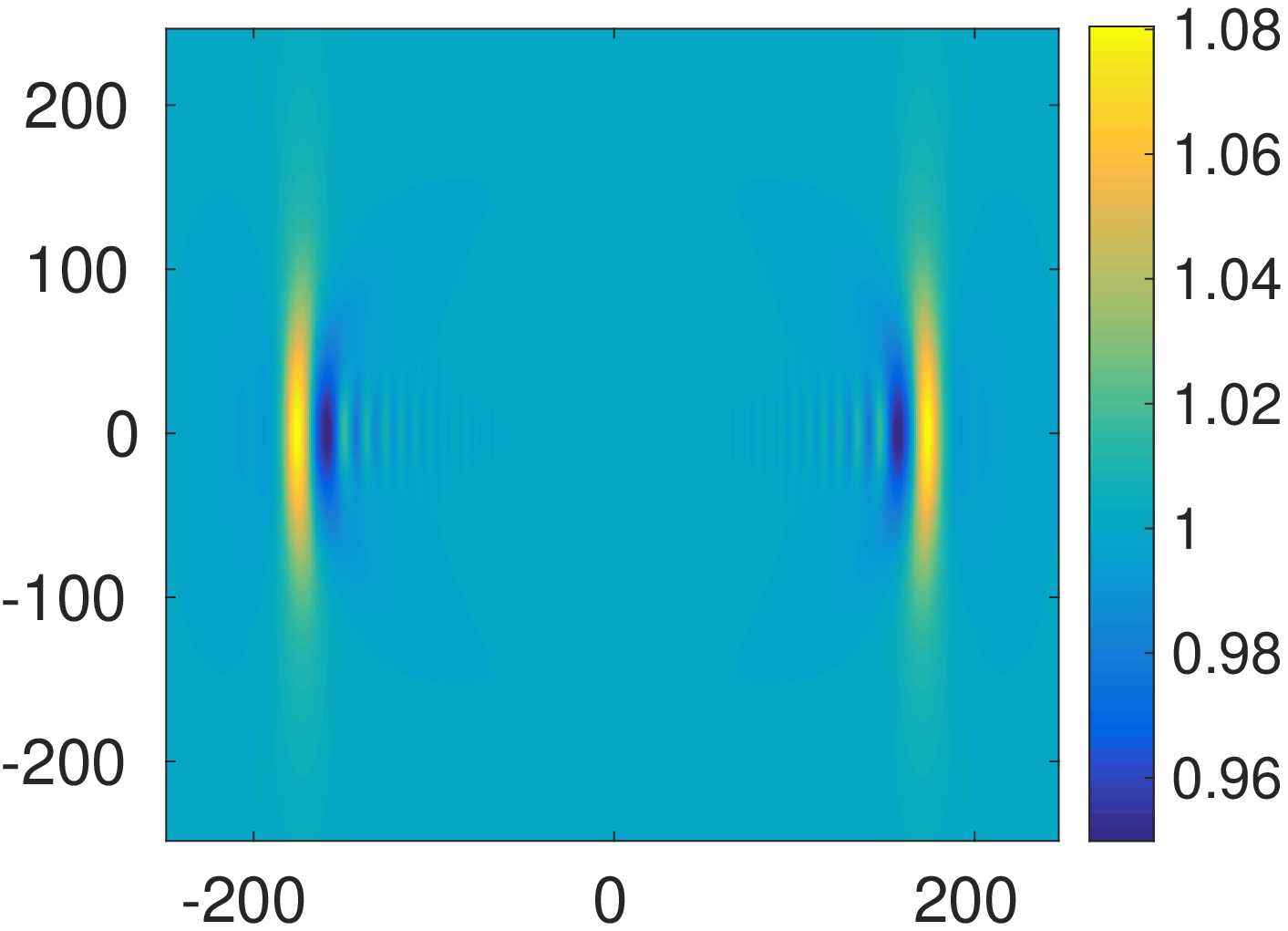}}
\caption{(Color online) Large-amplitude anti-dark lump solitons. Top panels show 
the evolution of this waveform for $\alpha=\frac{1}{2}$, $\epsilon=0.01$, $\gamma=1$, and $\beta=3$; 
it is observed that this structure evolves into a bent shape.  
Bottom panels depict the collision between two identical anti-dark lumps, for  
$\alpha=0.6$, $\epsilon=0.08$, $\gamma=0$, and $\beta=\frac{1}{2}$; the collision appears 
to be almost elastic although radiation develops in each lump after it.
} 
		\label{Fig5}
	\end{figure}

\begin{figure}[tbp]
		\subfloat[t=0]{\includegraphics[width=.26\textwidth]{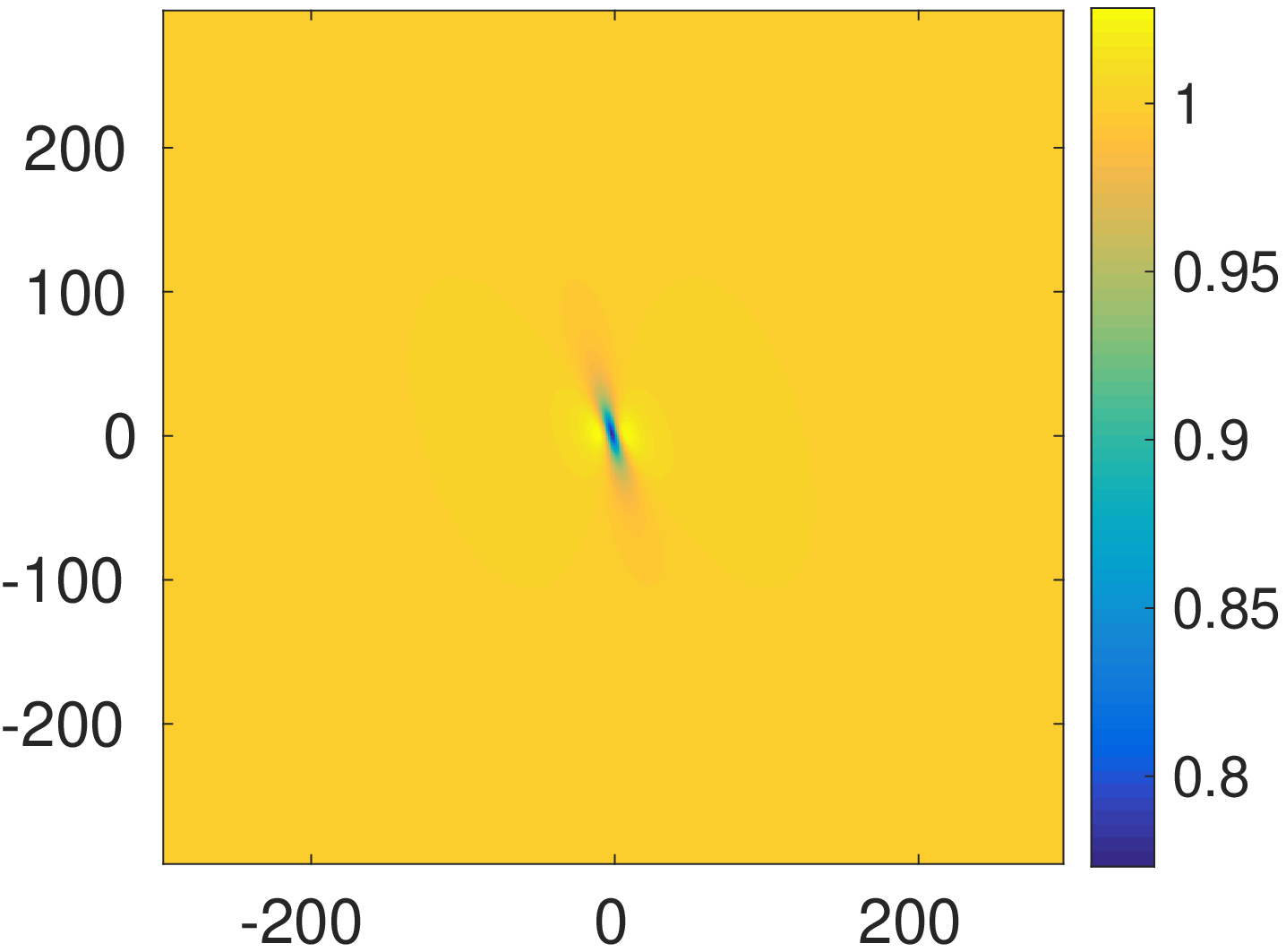}}\quad
		\subfloat[t=66]{\includegraphics[width=.26\textwidth]{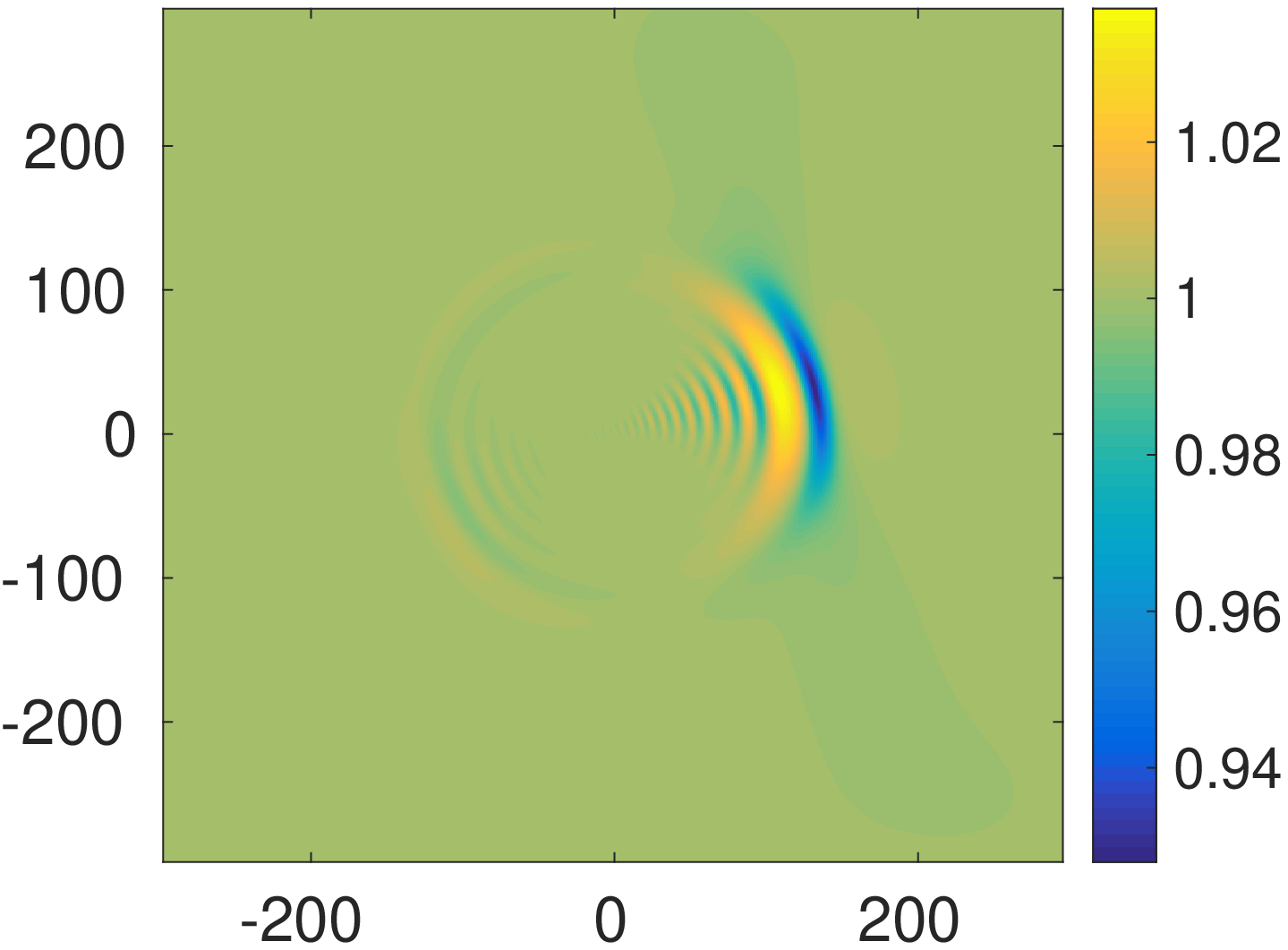}}\quad
		\subfloat[t=100]{\includegraphics[width=.26\textwidth]{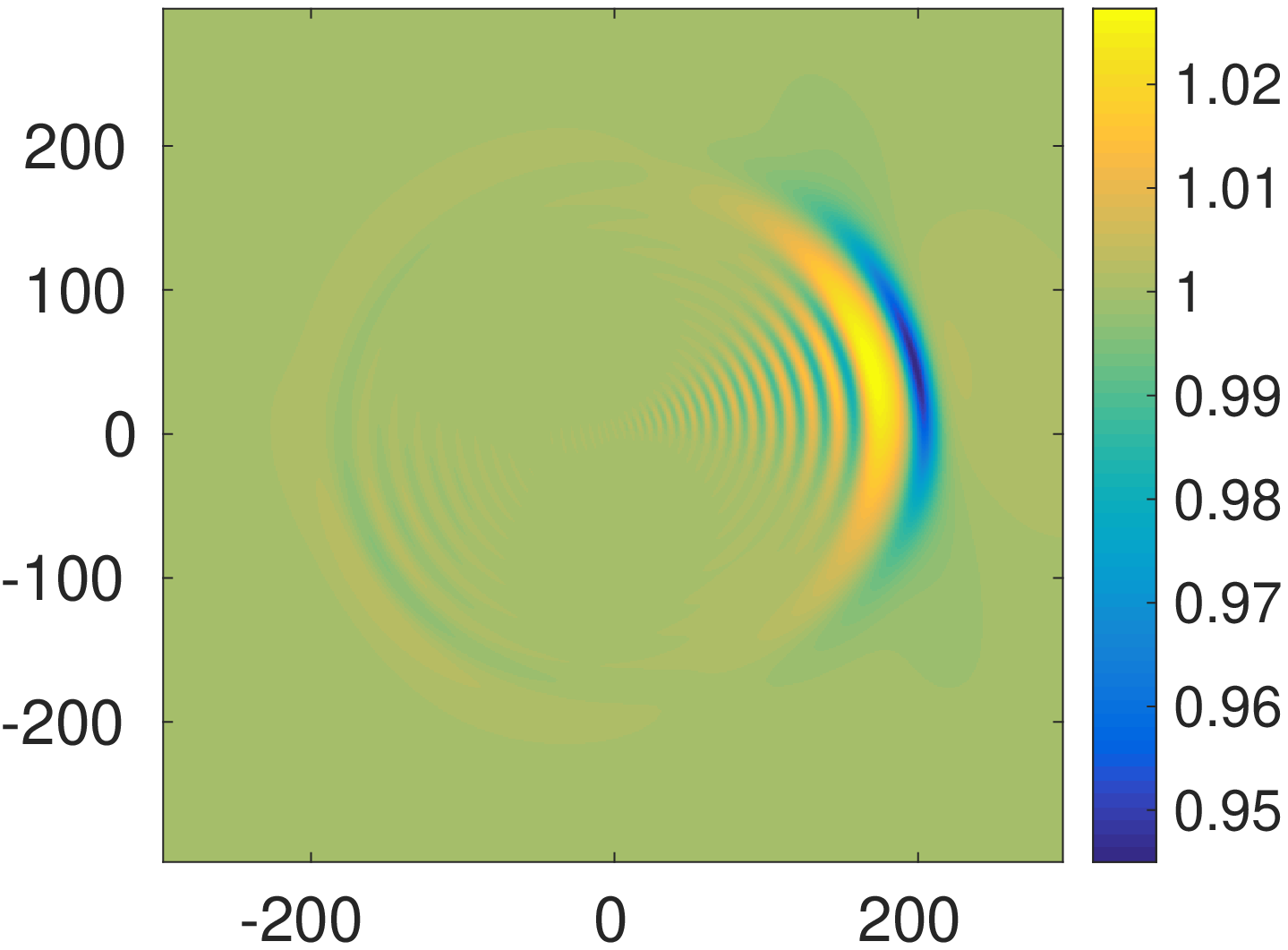}} \\
	\subfloat[t=0]{\includegraphics[width=.26\textwidth]{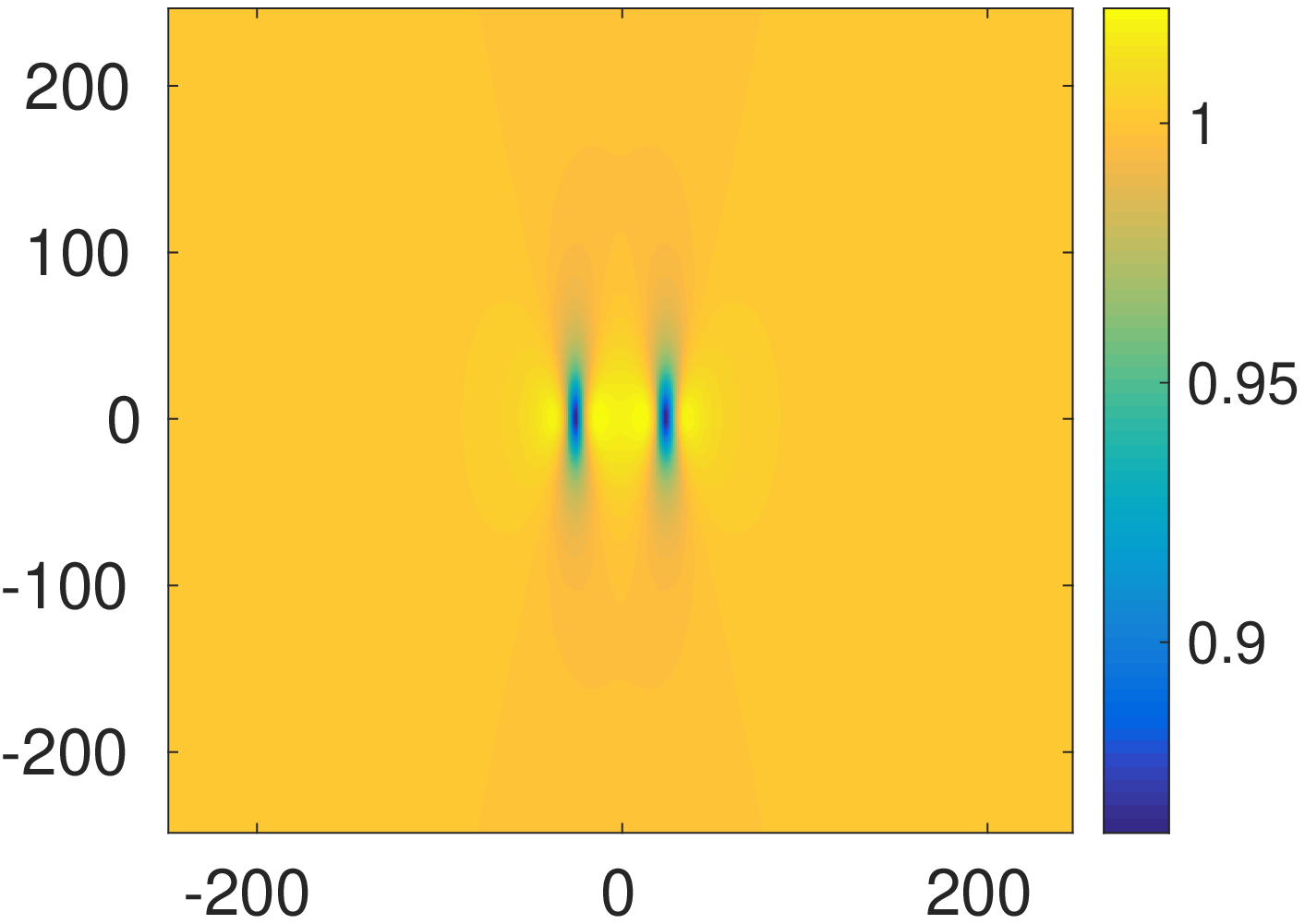}}\quad
		\subfloat[t=33]{\includegraphics[width=.26\textwidth]{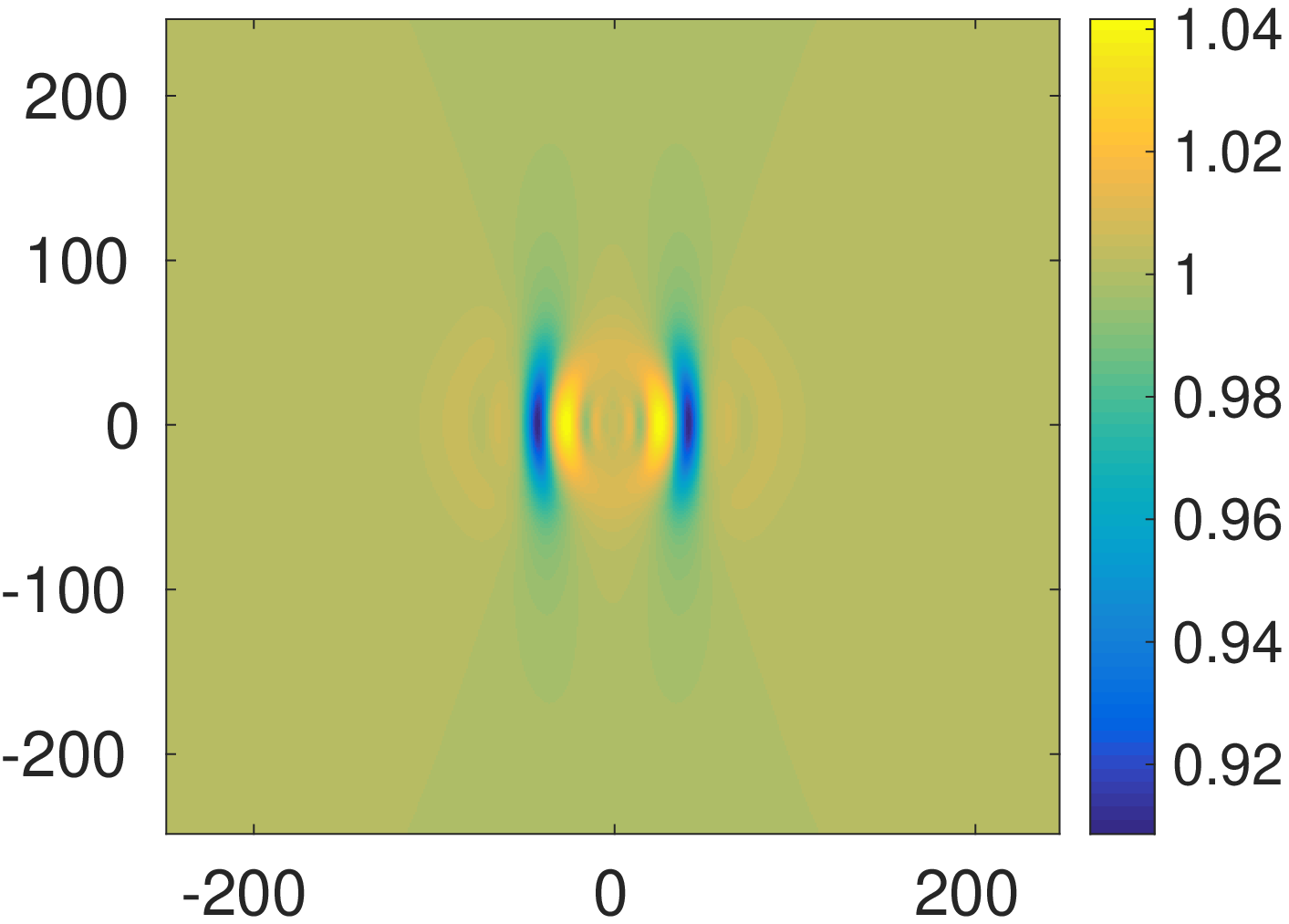}}\quad
	\subfloat[t=100]{\includegraphics[width=.26\textwidth]{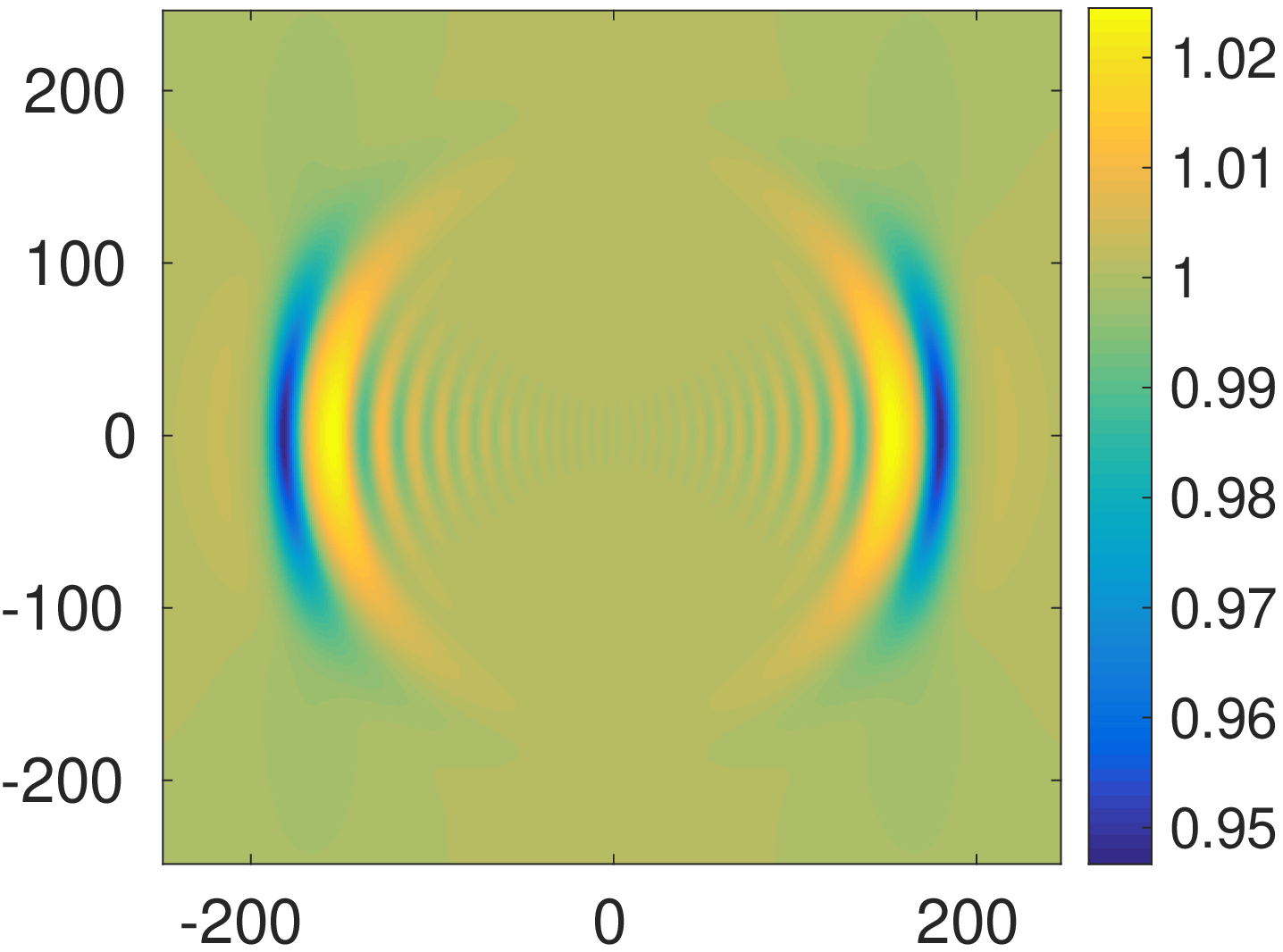}}
\caption{(Color online) Similar to Fig.~\ref{Fig5}, but for large-amplitude dark lump solitons. 
The observed behavior is similar to that of the anti-dark lump case. For the top panels, 
the parameter values are $\alpha=1$, $\epsilon=0.01$, $\gamma=1$, and $\beta=1.8$, 
while for the bottom panels $\alpha=1$, $\epsilon=0.08$, $\gamma=0$, and $\beta=\frac{1}{2}$.
} 
		\label{Fig6}
	\end{figure}

Next, we consider lumps of larger amplitudes. Figures~\ref{Fig5} and \ref{Fig6} 
depict the evolution and collisions of anti-dark and dark lump solitons, respectively. 
Regarding their evolution (top panels of Figs.~\ref{Fig5} and \ref{Fig6}), it is observed that  
although both structures remain localized, 
they spread and bend radially, emitting also radial radiation. 
Indeed, as seen in the more pronounced case of the dark lump of Fig.~\ref{Fig6}, 
the emitted radiation has the form of nearly concentric circles. 
The collision between anti-dark or dark lumps (bottom panels 
of Figs.~\ref{Fig5} and \ref{Fig6}, respectively) appears to be elastic; nevertheless, post 
collision dynamics again features the formation of (nearly) concentric
segments of circular rings, which 
appear to be more pronounced in the case of dark lumps.

\begin{figure}[!htbp]
		\subfloat[t=0]{\includegraphics[width=.26\textwidth]{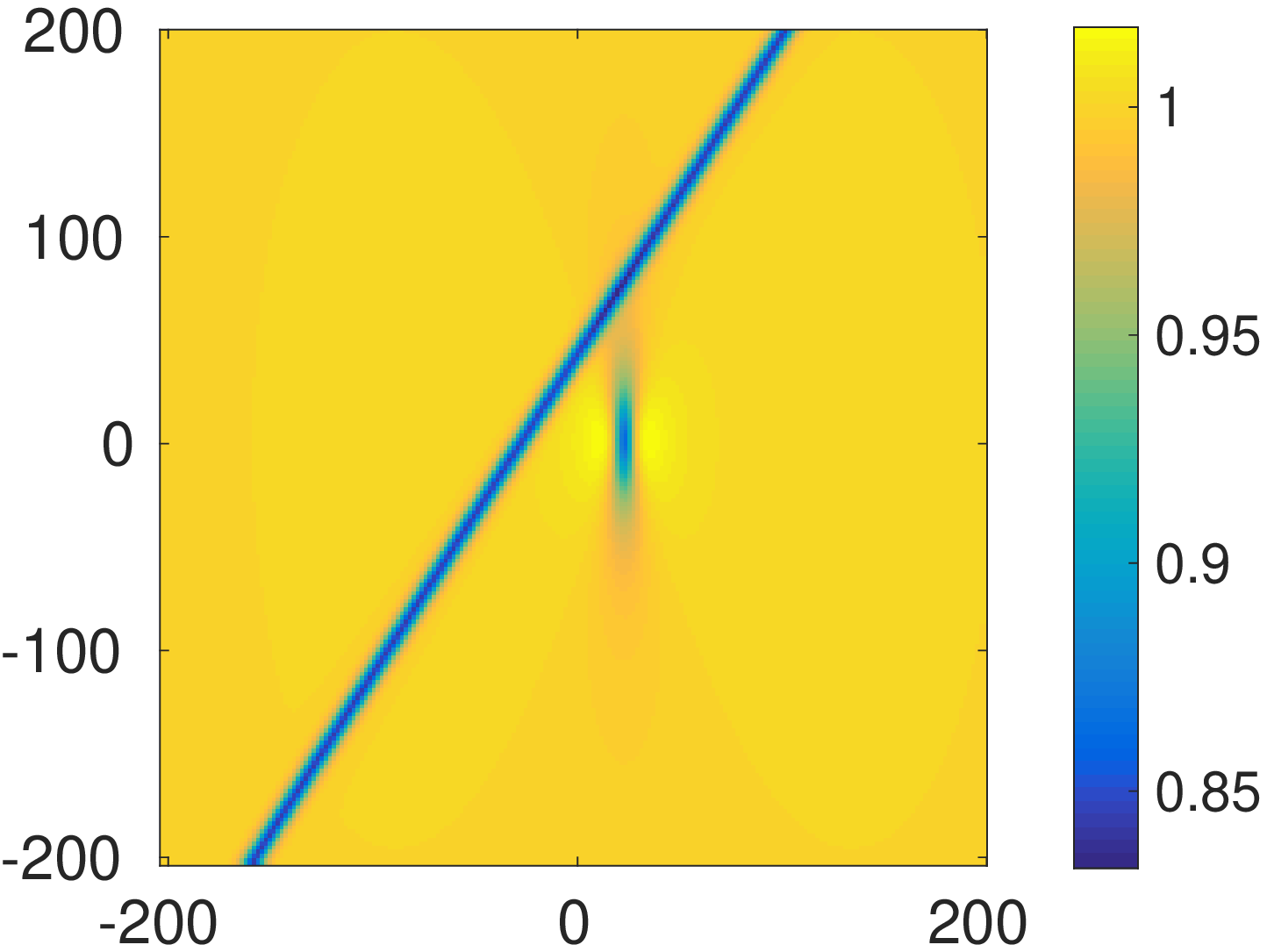}}\quad
		\subfloat[t=33]{\includegraphics[width=.26\textwidth]{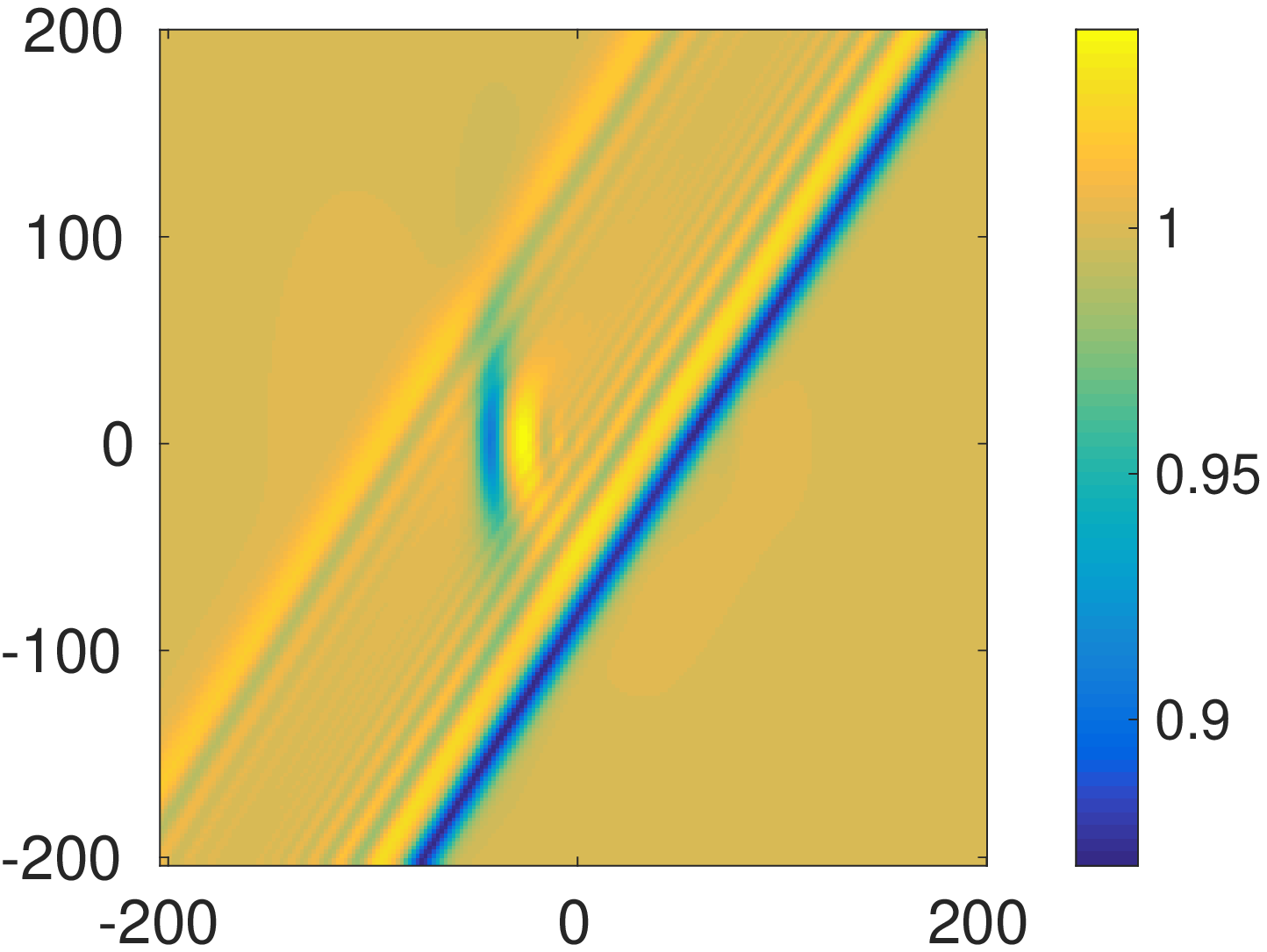}}\quad
		\subfloat[t=66]{\includegraphics[width=.26\textwidth]{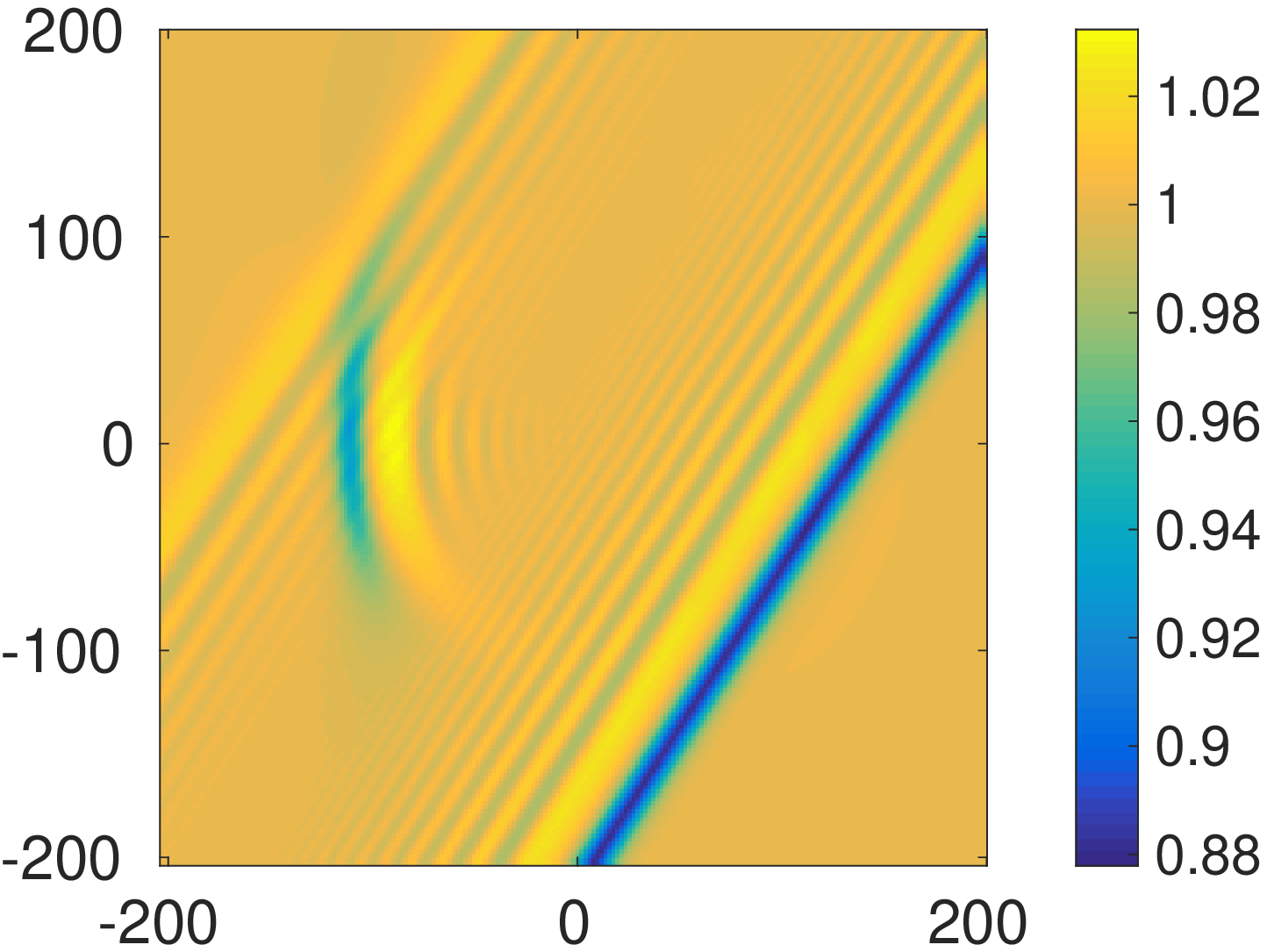}}\quad
		\subfloat[t=0]{\includegraphics[width=.26\textwidth]{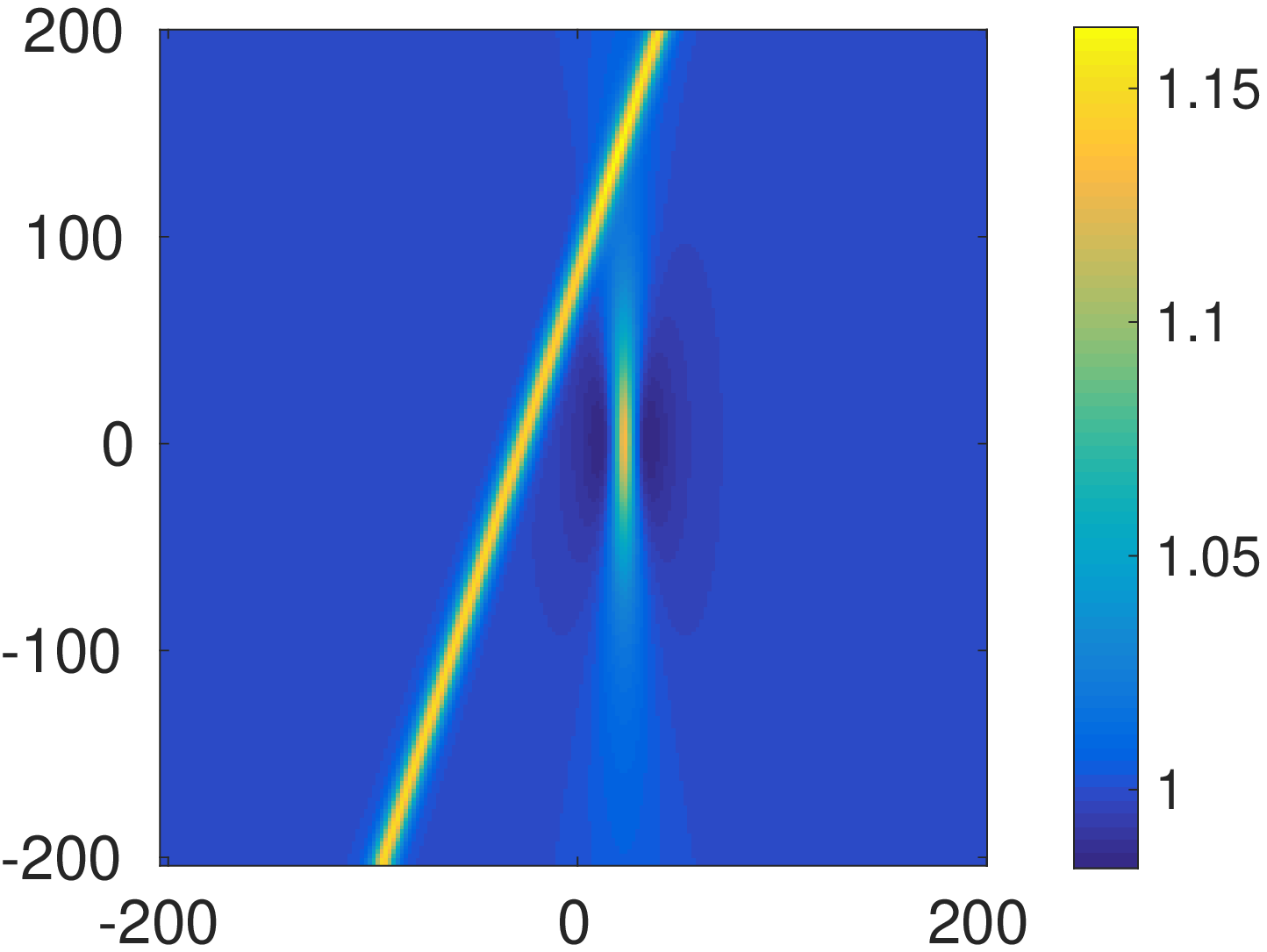}}\quad
		\subfloat[t=33]{\includegraphics[width=.26\textwidth]{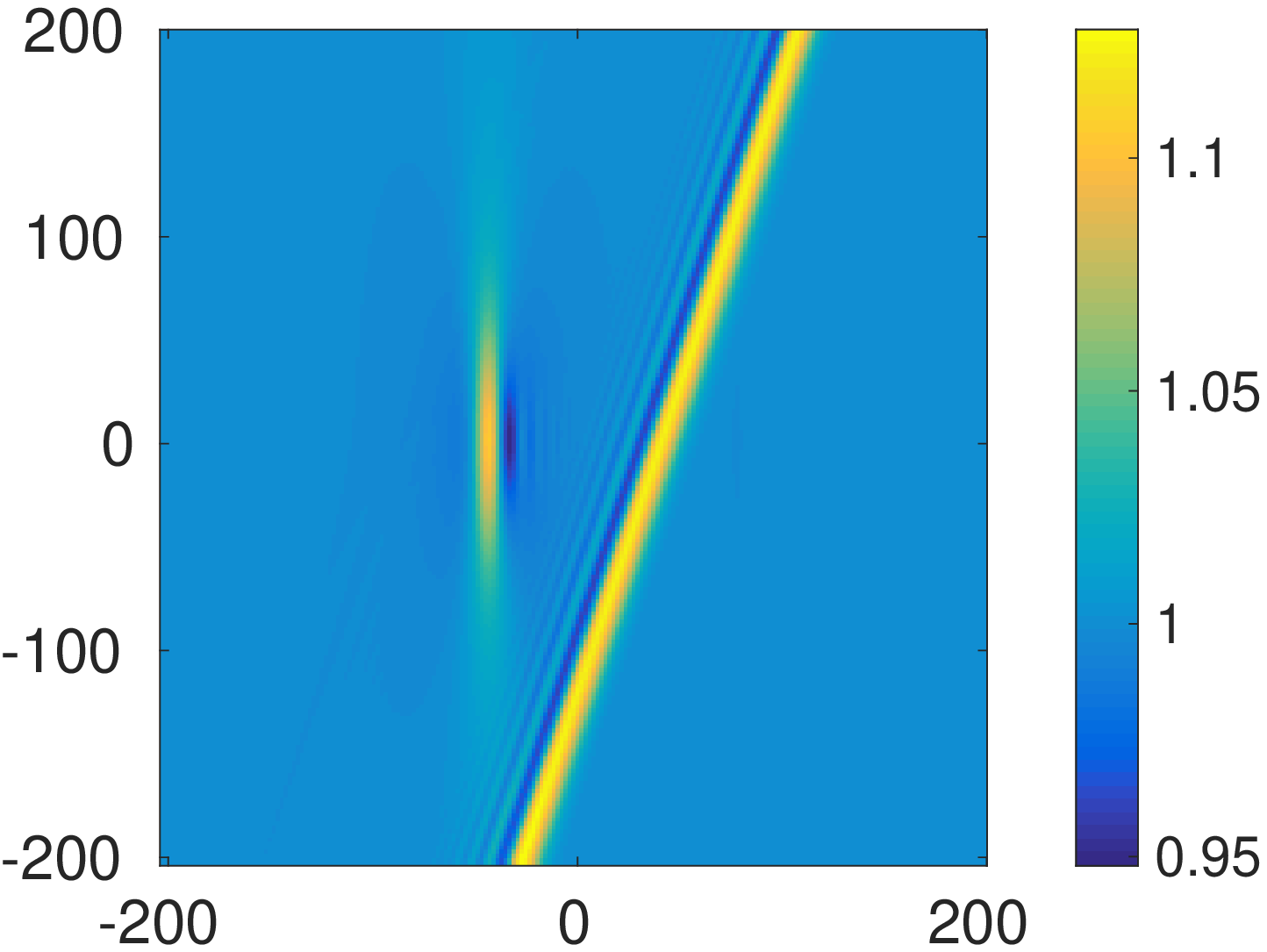}}\quad
		\subfloat[t=66]{\includegraphics[width=.26\textwidth]{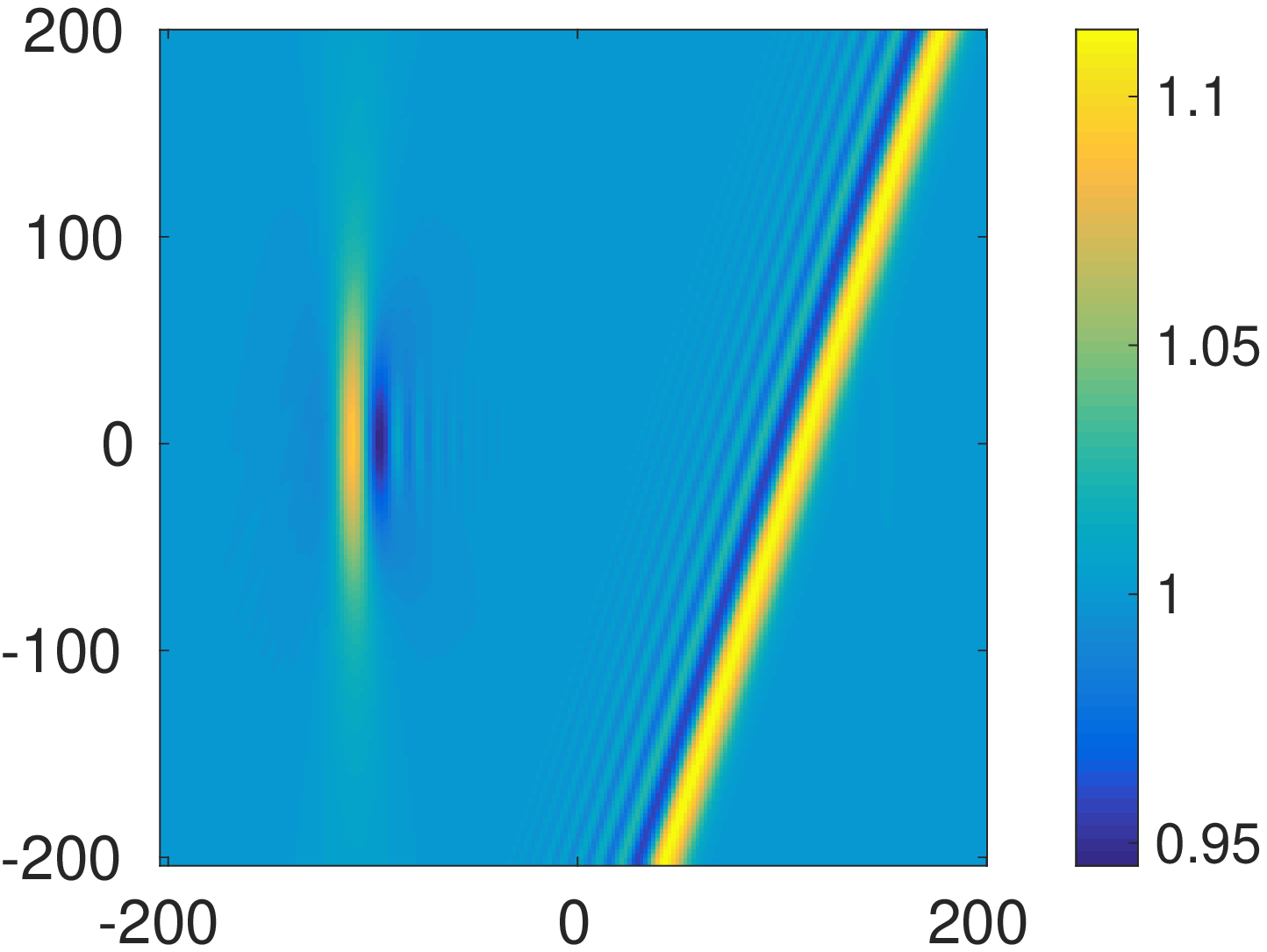}}\quad
		\caption{(Color online) Top panels: collision between large-amplitude dark line and lump solitons;  
parameters used: $\alpha=1$, $\epsilon=0.08$, $\gamma_{\text{Line}}=1.5$, $\beta_{\text{Line}}=1$, $\gamma_{\text{Lump}}=0$, and $\beta_{\text{Lump}}=\frac{1}{2}$.
Bottom panels: collisions between large-amplitude anti-dark line and lump solitons; parameters used: 
$\alpha=0.6$, $\epsilon=0.08$, $\gamma_{\text{Line}}=1.5$, $\beta_{\text{Line}}=1$, 
$\gamma_{\text{Lump}}=0$, and $\beta_{\text{Lump}}=\frac{1}{2}$.
		} \label{Fig8}
	\end{figure}

We have also performed simulations to study collisions between line and lump solitons. 
Pertinent results are depicted in Fig.~\ref{Fig8} for solitons of both types, dark (top panels) 
and anti-dark (bottom panels). In this case too, nearly elastic collision occurs in both cases, 
with the deformation of the lumps along the radial direction persisting as in the previously 
studied cases.
We note in passing that collisions between line dark solitons and 
anti-dark lumps (and vice versa) are not possible because solitons of the dark and the anti-dark 
type and of different dimensionality do not coexist for the same parameter values; 
such collisions may become possible only in the presence of higher-order effects that may 
facilitate the coexistence of such structures (see, e.g., Ref.~\cite{hec} where third-order 
dispersion supports solitons of different types and different dimensionality, which can undergo 
quasi-elastic head-on collisions).

\begin{figure}[!htbp]
		\subfloat[t=0]{\includegraphics[width=.26\textwidth]{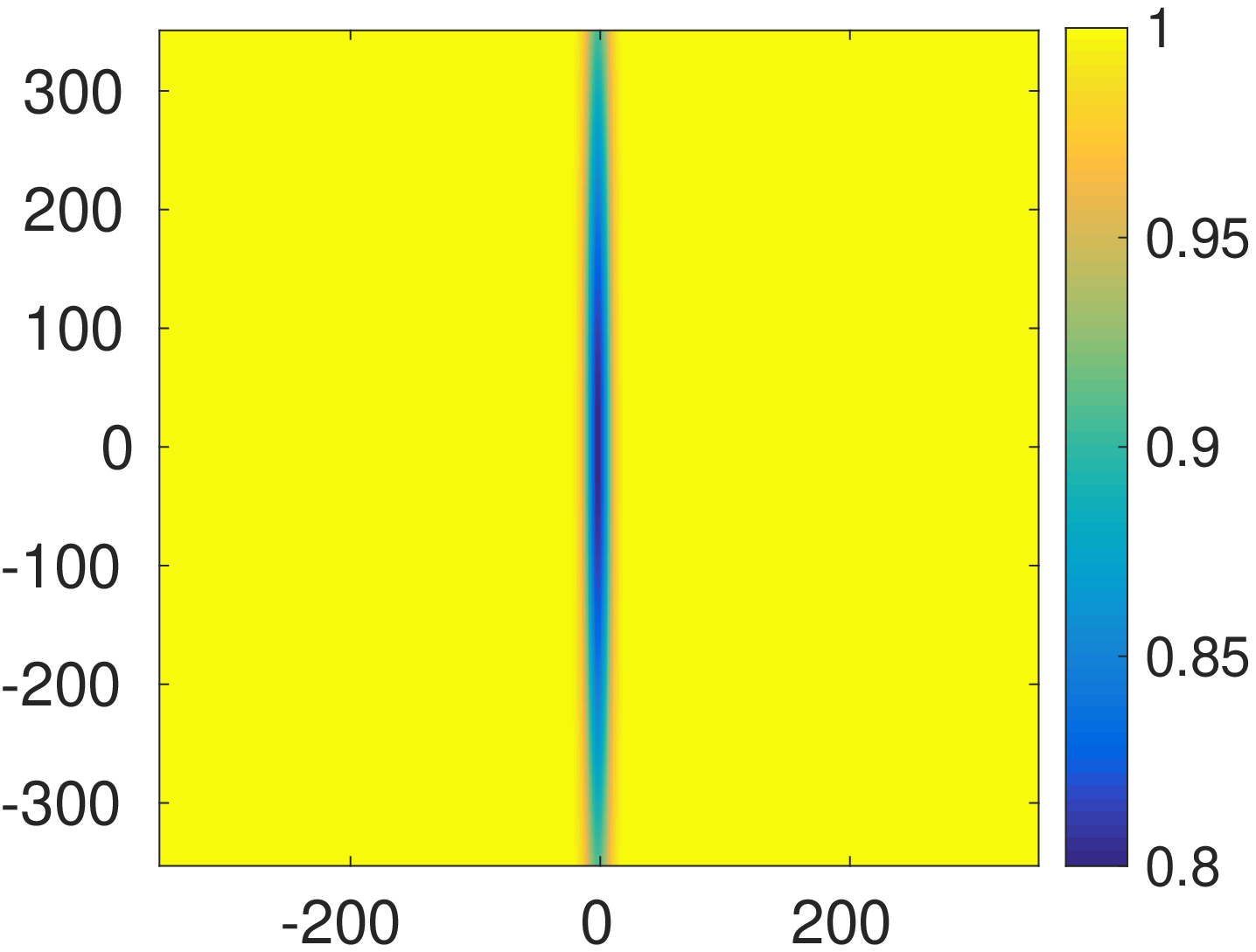}}\quad
		\subfloat[t=33]{\includegraphics[width=.26\textwidth]{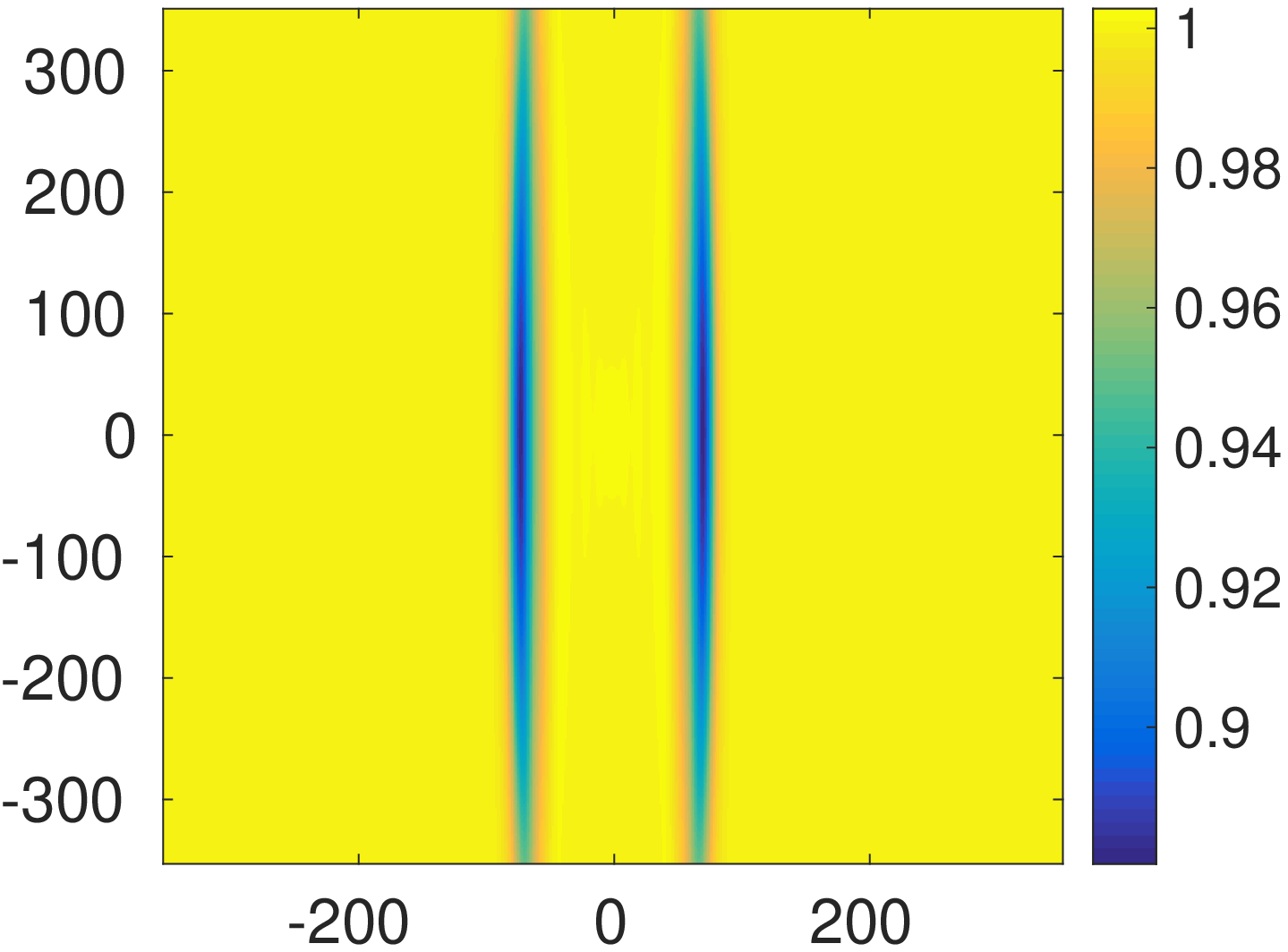}}\quad
		\subfloat[t=66]{\includegraphics[width=.26\textwidth]{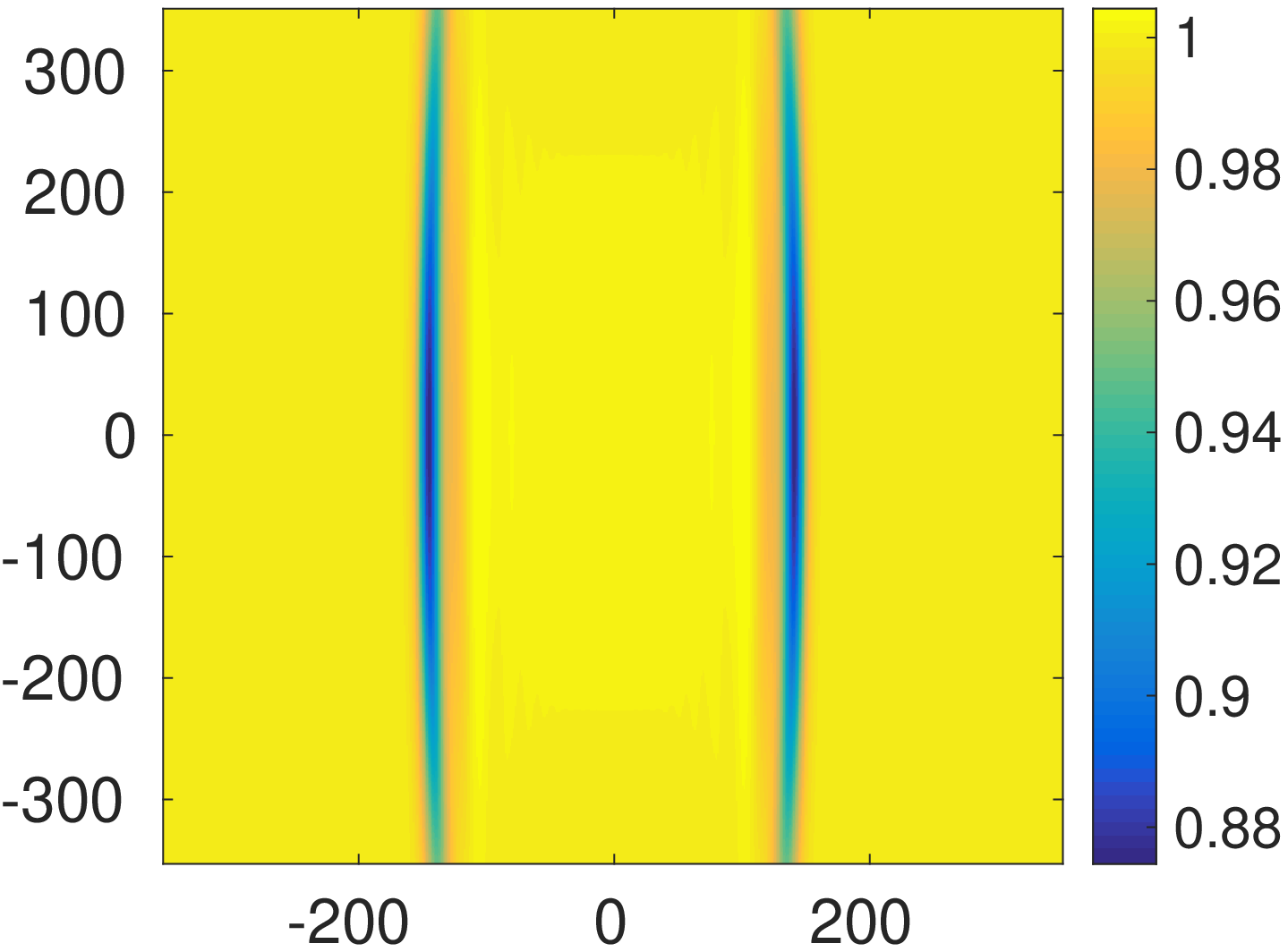}}\quad
		\subfloat[t=0]{\includegraphics[width=.26\textwidth]{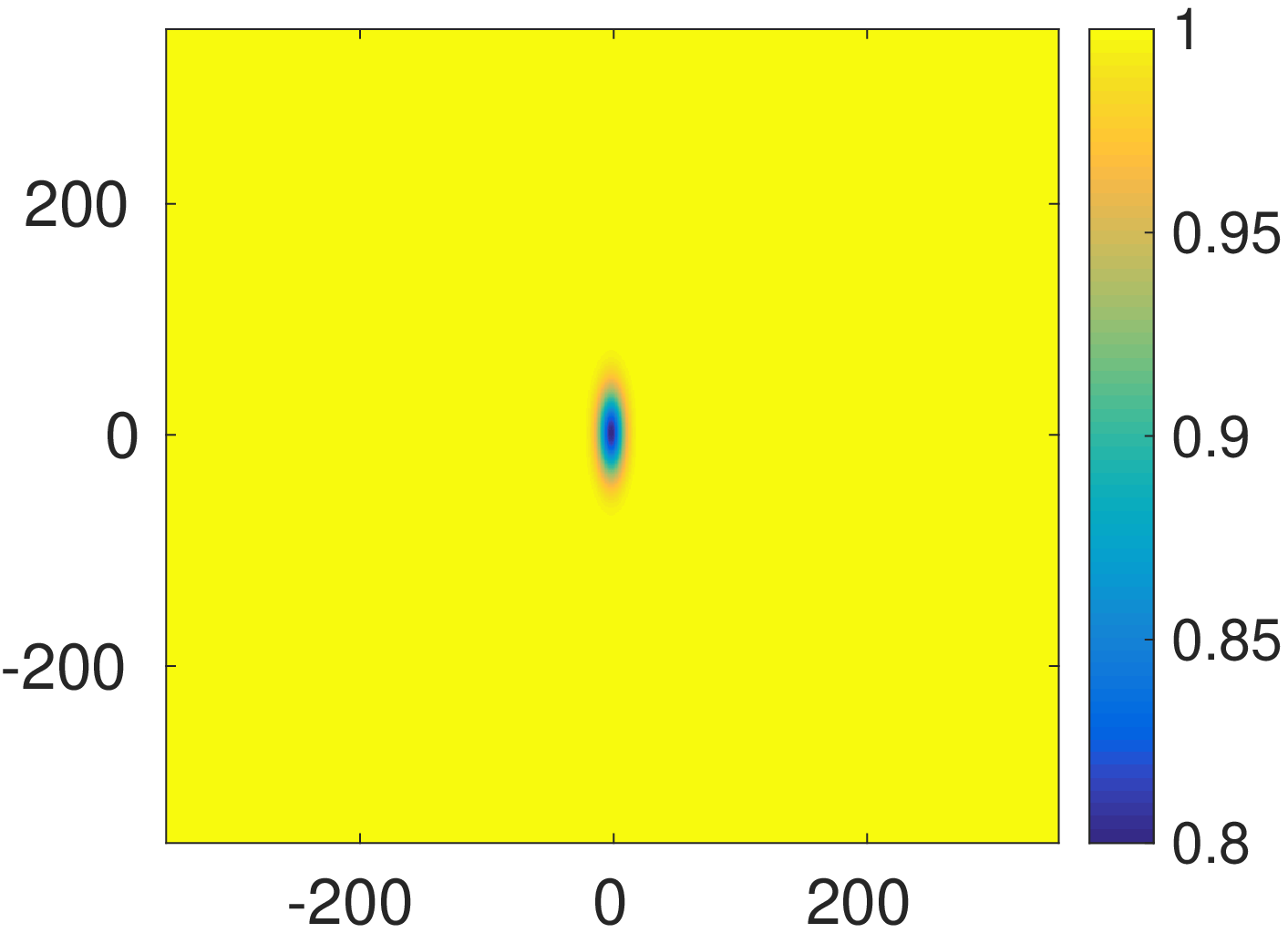}}\quad
		\subfloat[t=33]{\includegraphics[width=.26\textwidth]{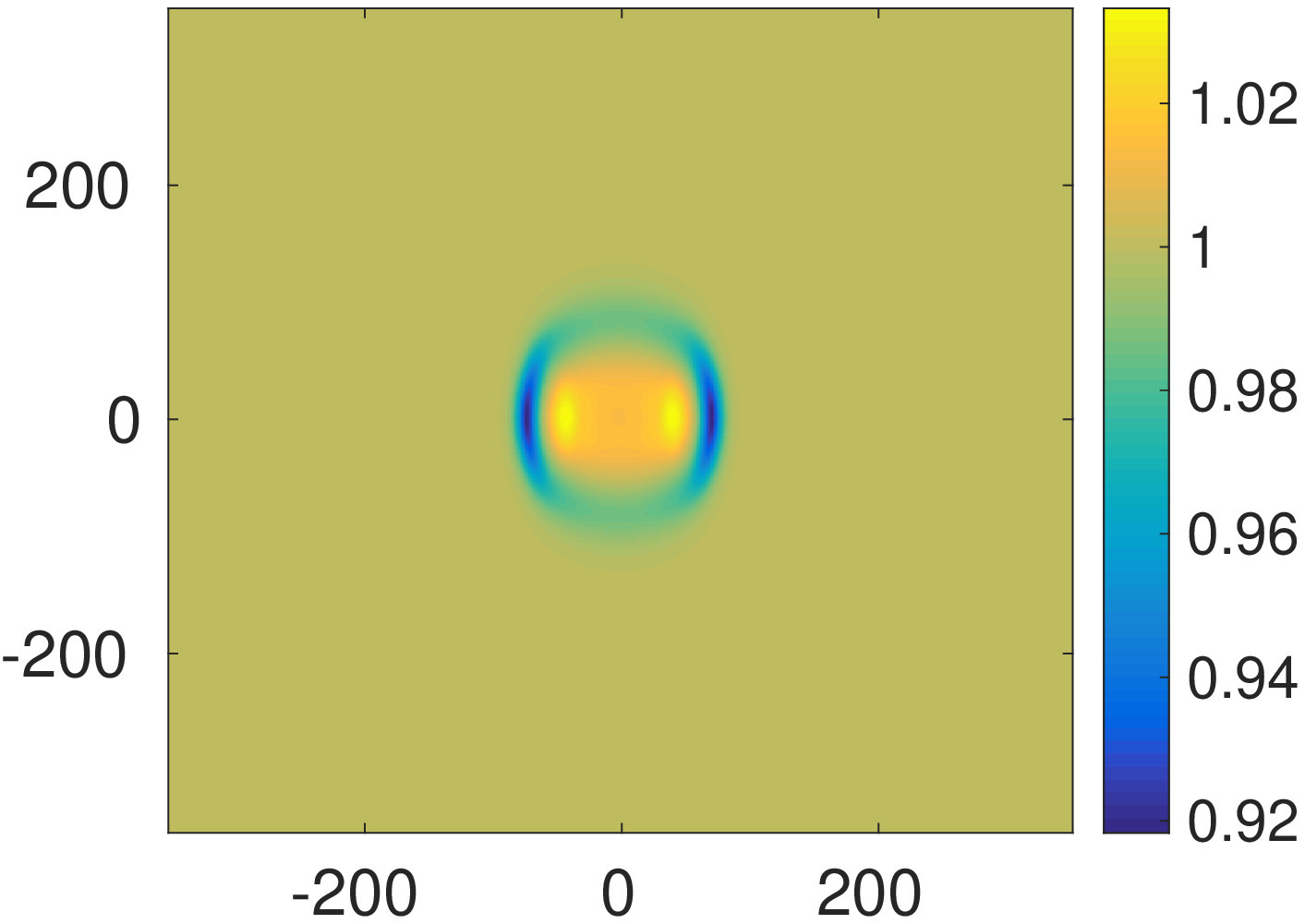}}\quad
		\subfloat[t=66]{\includegraphics[width=.26\textwidth]{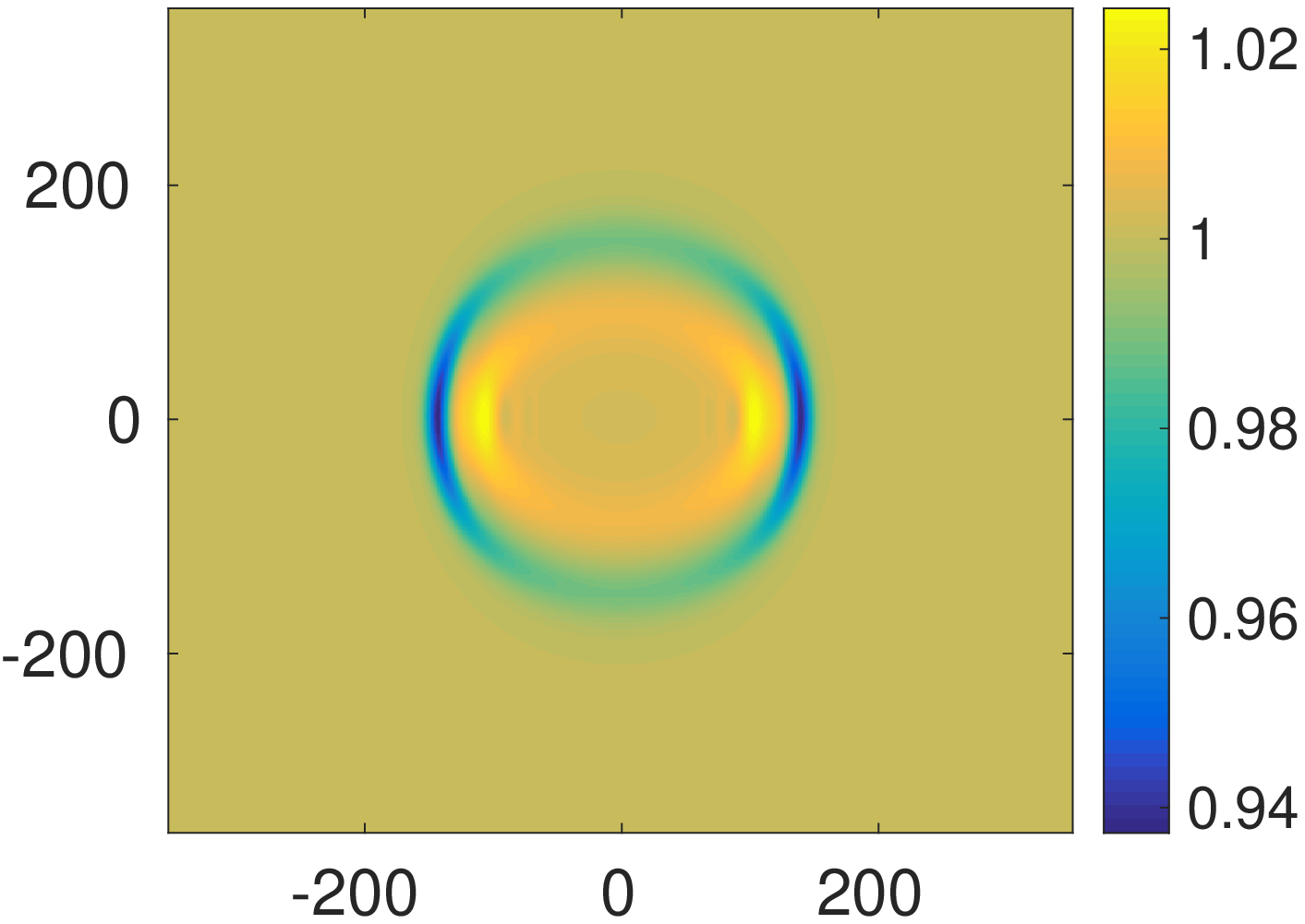}}\quad
		\subfloat[t=0]{\includegraphics[width=.26\textwidth]{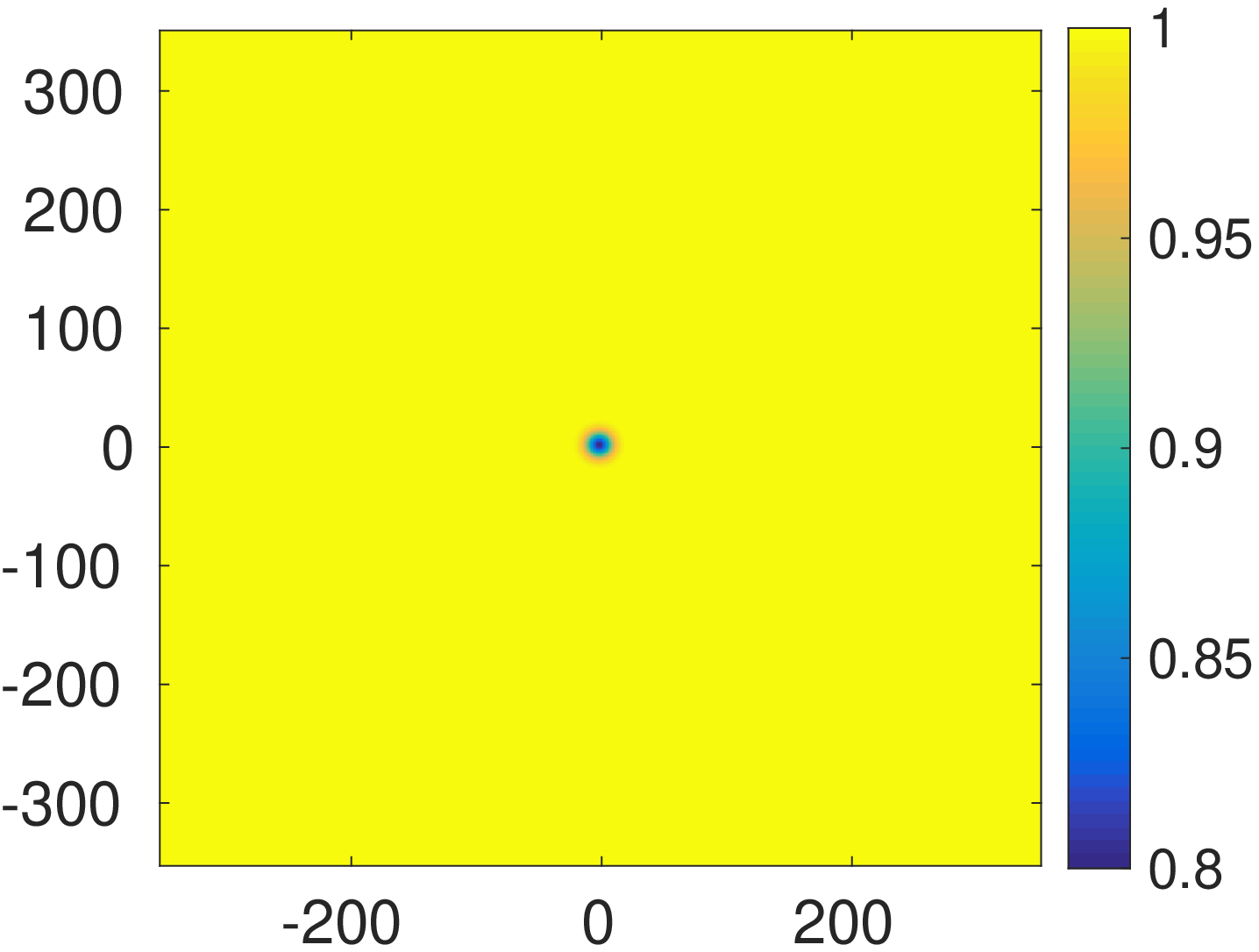}}\quad
		\subfloat[t=33]{\includegraphics[width=.26\textwidth]{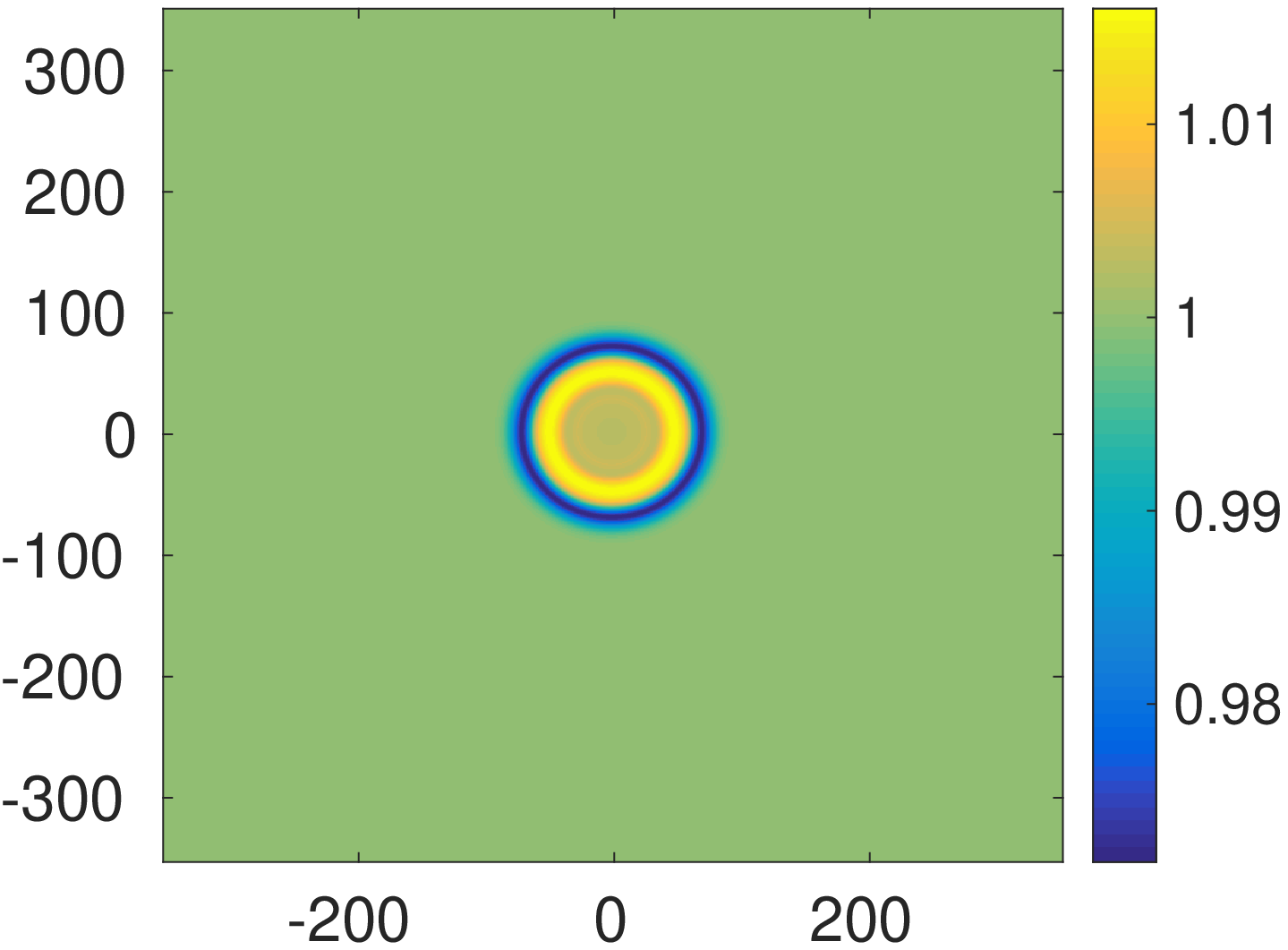}}\quad
		\subfloat[t=66]{\includegraphics[width=.26\textwidth]{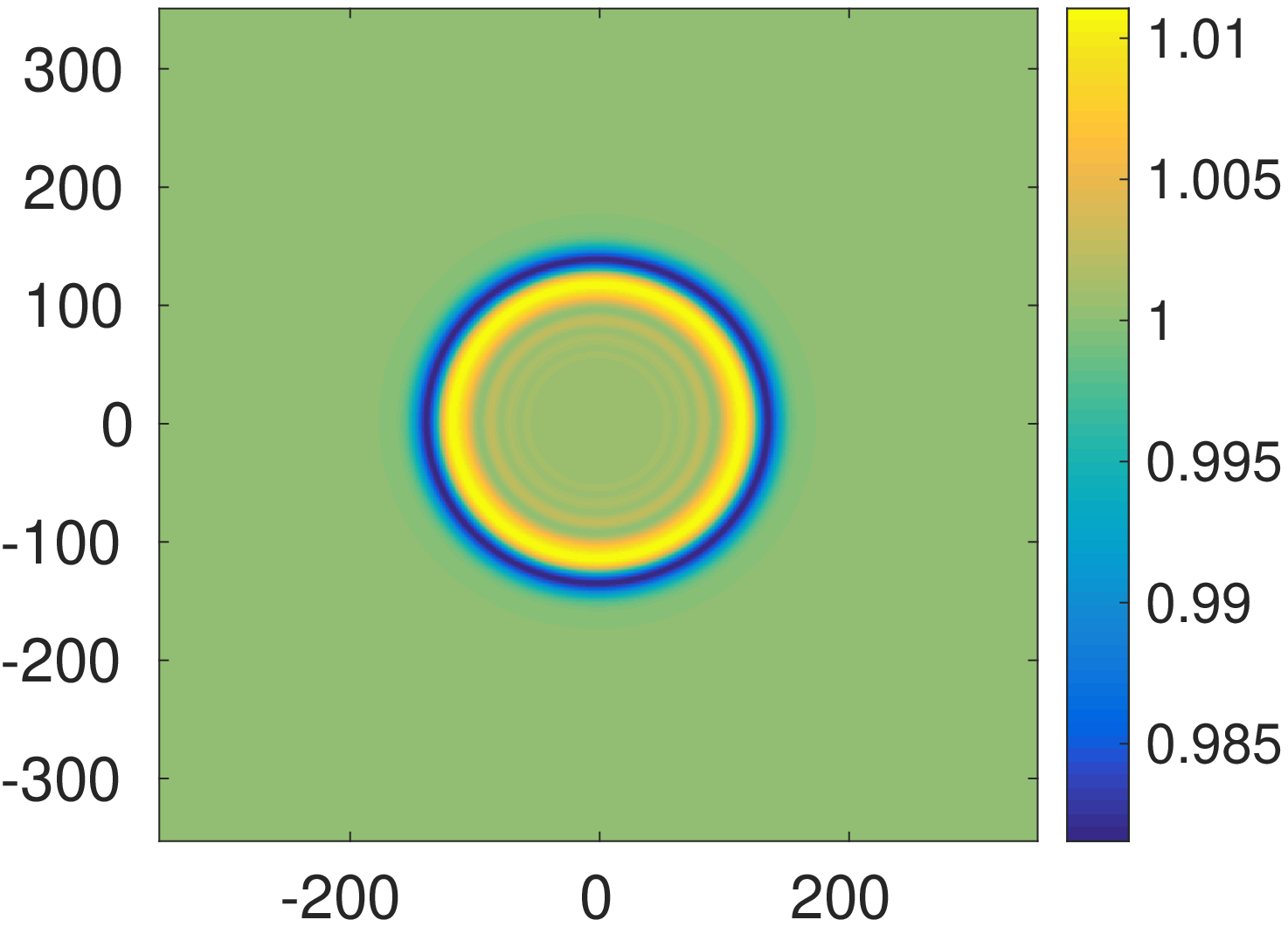}}\quad
		\caption{(Color online) Evolution of Gaussian-like pulses; 
the top panels depict the case of an extremely elongated (in the $y$-direction) such pulse, the
middle panels show the case of a slightly elongated Gaussian, while
the bottom panels 
depict the evolution of a completely symmetric such pulse. For detailed
initial condition parameters, see the text.
} 
\label{Fig12}
	\end{figure}

Lastly, we use generic Gaussian initial data on top of the background 
to investigate whether the resulting dynamics can share some qualitative features with 
the one corresponding to the approximate line and lump soliton solutions. To be exact, for the initial condition we place the Gaussian
\begin{equation}
u(x,y,0)=1+  A \exp \left[-\left(\frac{x}{\sigma_1 L_x}\right)^2 - \left(\frac{y}{\sigma_2 L_y}\right)^2\right]
\end{equation}

on top of the background, where $A=-0.2$, $L_x$, $L_y$ are the bounds of the computational domain and $\sigma_1$, $\sigma_2$ control the length and width. We have fixed $L_x=L_y=500$ and $\sigma_1=0.02$. Pertinent 
results are shown in Fig.~\ref{Fig12}, for a Gaussian
very stretched in the 
vertical direction (top panels, $\sigma_2=0.8$), for a slightly stretched Gaussian (middle panels, $\sigma_2=0.07$), and 
for a completely symmetric Gaussian (bottom panels, $\sigma_2=0.02$). 

It is observed that, in the first 
case, the extended Gaussian splits into two symmetric waveforms reminiscent
of a pair 
of dark line solitons moving in opposite directions. Here, the initial stripe-like 
structure, is similar to a line dark soliton but without a phase jump [cf. Eq.~(\ref{29})]. 
To attain the correct phase profile suggested by the energetically preferable 
approximate line dark soliton, and still preserve the phase structure at infinity, 
the initial Gaussian has to split to two oppositely moving stripes. 

On the other hand, in the case where the initial condition has the form of a slightly 
extended Gaussian (cf. middle panels of Fig.~\ref{Fig12}), the form of the initial data 
is closer to that of a dark lump soliton (rather than that of a line soliton). In this case too, 
due to not having the proper phase structure --and decay at infinity which is now exponential 
rather than algebraic-- the initial data reorganizes itself into a structure which is 
reminiscent to a superposition of two bent lumps.
Here, one should observe the resemblance of this dynamics with the outcome of 
the collision between two dark lumps of Fig.~\ref{Fig6}. Finally,
when the initial 
Gaussian is completely symmetric, the resulting structure evolves towards
an almost 
perfect ring-like structure resembling an expanding ring soliton.

\section{Conclusions \& Future Challenges}

In this work, we have studied the defocusing Camassa-Holm--Nonlinear Schr{\"o}dinger (CH-NLS) 
equation. We have shown that this model possesses a stable continuous-wave (cw) solution, on 
top of which small-amplitude soliton solutions can be supported. Our analytical approach 
was based on asymptotic multiscale expansion methods, which allowed us to reduce the CH-NLS 
model to a Kadomtsev-Petviashvili (KP) equation. Both versions, namely the KP-I and KP-II, 
were found to be possible, depending on the sign of a characteristic parameter.

The reduction to the KP model allowed us to construct approximate soliton solutions,  
both line solitons and lumps, and either of the dark or of the anti-dark type, of the 
original CH-NLS model. Domains of existence of all these structures, as well as their 
dynamics by means of direct numerical simulations, were investigated. We found that 
line and lump solitons do persist in the original model, but as their amplitude is 
increased, they undergo deformations, i.e., bending and a radial expansion, which 
may form other, ring-shaped, structures. We also studied head-on collisions between 
line solitons, between lumps, as well as between line solitons and lumps, and found 
that they are almost elastic (although less so as the amplitude of the
structures increases). In our simulations, we have also used generic Gaussian 
initial data, the dynamics of which were found to follow qualitative features of the 
approximate soliton solutions' dynamics.

There are many interesting topics for future studies. First of all, it would be 
interesting to study the transverse dynamics of large-amplitude dark solitons and 
investigate their instability, and the concomitant generation of vortices 
(similarly to the traditional 2D defocusing NLS model \cite{pel1,pel2}). 
The study of other quasi-2D or purely 2D structures, such as the ring dark or 
anti-dark solitons (which were already identified in our simulations) and vortices, 
respectively, constitute still other themes of particular interest, as also
highlighted by select ones among our numerical computations (e.g., the
case of the radially symmetric Gaussian initial condition). Relevant studies 
are in progress and pertinent results will be reported in future publications.\\
\\
\indent
This material is based upon work supported by the National Science Foundation under Grant No. PHY-1602994.

\end{document}